\documentclass[usenatbib]{mnras}
\usepackage{newtxtext,newtxmath}

\usepackage[T1]{fontenc}
\usepackage{ae,aecompl}
\usepackage{pdflscape}

\usepackage{graphicx}	
\usepackage{amsmath}	
\usepackage{amssymb}	
\usepackage{booktabs}
\usepackage{threeparttable}
\usepackage[caption = false]{subfig}

\newcommand{\EQ}[1] {Equation~(\ref{#1})}
\newcommand{\SEC}[1] {Section~\ref{#1}}
\newcommand{\APP}[1] {Appendix~\ref{#1}}
\newcommand{\FIG}[1] {Figure~\ref{#1}}
\newcommand{\TAB}[1] {Table~\ref{#1}}

\newcommand{\VEC}[1] {{\boldsymbol{{ #1}}}}
\newcommand{\MX}[1] {{\mathbf{{ #1}}}}

\newcommand{\D} {\mathrm{d}}

\def\DM{\mathrm{DM}}
\def\DMm{\DM_{\rm MW}}
\def\DMd{\DM_{\rm halo}}
\def\DMe{\DM_{\rm E}}
\def\DMi{\DM_{\rm IGM}}
\def\DMh{\DM_{\rm host}}
\def\DMho{\DM_{\rm host,0}}
\def\DMs{\DM_{\rm src}}

\def\Ha{\mathrm{H}\alpha}

\def\fr{f_r}
\def\fz{f_z}
\def\fd{f_{\rm \cal D}}
\def\fs{f_\mathrm{s}}
\def\fe{f_\mathrm{\epsilon}}

\def\cmpc{\mathrm{cm}^{-3}\, \mathrm{pc}}
\def\kpc{\mathrm{kpc}}
\def\ergs{\mathrm{erg}\, \mathrm{s}^{-1}}

\title[Normalised FRB luminosity function]{On the normalised FRB luminosity
function}

\author[R. Luo et al.]{
Rui Luo $^{1,2}$, Kejia Lee\thanks{E-mail: kjlee@pku.edu.cn} $^{1,3}$, Duncan R. Lorimer $^{4,5}$ and Bing Zhang $^{1,2,6}$\\
$^{1}$ Kavli Institute for Astronomy and Astrophysics, Peking University, 
Beijing 100871, China \\
$^{2}$ Department of Astronomy, School of Physics, Peking University, Beijing 
100871, China\\
$^{3}$ National Astronomical Observatories, Chinese Academy of Sciences, 
Beijing 100012, China\\
$^{4}$ Department of Physics and Astronomy, West Virginia University, Morgantown, 
WV 26506, USA \\
$^{5}$ Center for Gravitational Waves and Cosmology, West Virginia University, 
Chestnut Ridge Research Building, Morgantown, WV 26505, USA \\
$^{6}$ Department of Physics and Astronomy, University of Nevada, Las Vegas, NV 89154, USA
}

\date{Accepted XXX. Received YYY; in original form ZZZ}

\pubyear{2018}
\begin{document}
\label{firstpage}
\pagerange{\pageref{firstpage}--\pageref{lastpage}}
\maketitle

\begin{abstract} Thirty-three fast radio bursts (FRBs) had been detected by March 2018. Although the sample size is still limited, meaningful statistical studies can already be carried out. The normalised luminosity
function places important constraints on the intrinsic power output, sheds light 
on the origin(s) of FRBs, and can guide future observations.  In this paper, we 
measure the normalised luminosity function of FRBs. Using Bayesian 
statistics, we can naturally account for a variety of factors such as
receiver noise temperature, bandwidth, and source selection
criteria. We can also include astronomical systematics, such as host galaxy 
dispersion measure, FRB local dispersion measure, galaxy evolution, geometric 
projection effects, and Galactic halo contribution. Assuming a Schechter  
luminosity function, we show that the isotropic luminosities of FRBs have a power-law distribution that
covers approximately three orders of magnitude, with a power-law 
index ranging from $-1.8$ to $-1.2$ and a cut off $\sim 2\times 10^{44}\,\ergs$. 
By using different galaxy models and well-established Bayesian marginalisation techniques, we show 
that our conclusions are robust against unknowns, such as the electron densities 
in the Milky Way halo and the FRB environment, host galaxy morphology, and telescope beam 
response.
\end{abstract}




\section{Introduction}

Fast Radio Bursts (FRBs) are a new type of radio transients,
and remain unexplained. The bursts last for a few milliseconds,
and show dispersive signatures with peak flux densities
ranging from 0.3~Jy to about 100 Jy. The first one (FRB~010724,
\citealt{Lorimer07Sci}) was detected serendipitously in the archival
data of pulsar survey for Small Magellanic Cloud using the
Parkes telescope \citep{MFL06}.  Shortly after that, a growing
number of FRBs were discovered with Parkes at 1.4
GHz, both in the archival data \citep{Keane12MN,Thornton13Sci, Burke-Spolaor14ApJ}
and from the real-time searches \citep{Ravi15ApJ,Petroff15MN, Keane16Nat,
Ravi16Sci, Petroff17MN, Bhandari18MN}. FRBs were
also detected by Arecibo \citep{Spitler14ApJ}, Green Bank Telescope \citep{Masui15Nat}, UTMOST \citep{Caleb17MN, Farah18MN} and ASKAP \citep{Bannister17ApJ}. At the time of 
writing this paper, the total number of the reported detections adds up to 33.

FRBs are natural celestial probes with a broad range of astrophysical
applications. For example, it has been proposed that FRBs could be used to test 
the Einstein's equivalence principle \citep{Wei15PRL,
Tingay16ApJ, Zhang16arXiv}, to constrain
the rest mass of photons \citep{Wu16ApJ,
Bonetti16PLB, Bonetti17PLB, Shao17PRD}, 
to detect the baryon contents in the Universe
\citep{McQuinn14ApJ}, to probe the cosmological matter 
distribution \citep{Masui15PRL}, to study the evolution of
intergalactic medium (IGM) \citep{Zheng14ApJ} and constrain the dark-energy
equation of states \citep{Zhou14PRD, Gao14ApJ}.

The origins of FRBs, however, remain mysterious and subject to an intensive 
debate.  Here, we list several proposals in the literature in chronological 
order: (1) radio pulses from black hole evaporative explosions \citep{Rees77Nat}; 
(2) superconducting cosmic strings \citep{CSV12, CSS12, Yu14JCAP}; (3) flaring 
magnetars \citep{PP10, PP13arXiv} or stars \citep{Loeb14MN}; (4) mergers of 
white dwarfs \citep{Kashiyama13ApJ}; (5) mergers of double neutron stars 
\citep{Totani2013PASJ, Wang16ApJ}; (6) collapses of neutron stars
into black holes \citep{FR14A&A, Zhang14ApJ}; (7) synchrotron masers 
\citep{Lyubarsky14MN, Ghisellini2017MN, Lu18MN}; (8) binary model of white dwarf 
and black hole \citep{GDL16};
(9) super-giant pulses from pulsars \citep{Cordes16MN, Connor16MN} ; (10) radio 
emission from soft gamma-ray repeaters \citep{Pen15ApJ, Katz16ApJ}; (11) axion 
stars \citep{Iwazaki15PRD}; (12) quark nova  \citep{Shand16RAA}; (13) mergers of 
charged black holes \citep{Zhang16ApJL, Liu16ApJ}; (14) collisions between 
pulsar and asteroids \citep{Geng15ApJ, Dai16ApJ}; (15) relativistic jet -- cloud 
interactions \citep{Romero16PRD, Vieyro17A&A};
(16) births of millisecond magnetars \citep{Metzger17ApJ};
(17) `cosmic comb', i.e. magnetosphere -- environment 
interactions\citep{Zhang17ApJ, Zhang18ApJ};  (18) accretion of black holes 
\citep{Katz17MN}; (19) star-quakes of compact stars \citep{Wang18ApJ}.

To understand the mechanisms of FRBs, the host galaxy information
is crucial. At this stage, only the repeating FRB, FRB~121102, had
the reliable identification of host galaxy \citep{Spitler16Nat,
Scholz16ApJ}.  \citet{Chatterjee17Nat} measured its precise position
using \emph{Karl G. Jansky Very Large Array}. The optical counterpart
was identified as a dwarf galaxy at the redshift of $z=0.193$
\citep{Tendulkar17ApJ}. However, we should be cautious in drawing
general remarks on the FRB environment, due to  unknown links between
repeating and non-repeating FRBs. Statistical analyses are needed
to quantify the properties of FRBs as an integrated population.

The normalised luminosity function, i.e. the probability density
function (PDF) of FRB luminosities, is one particularly important
statistics for the FRB intrinsic power output. The computation of
the luminosity functions requires not only FRB flux and distances,
but also a detailed account of any biases in the sample. For example,
without the counterpart identifications, the FRB distances are
usually estimated via the dispersion measure (DM). The estimated FRB
distance and luminosity are affected by the uncertainties in the
DM modelling. It is absolutely necessary to account for these effects
in inferring the luminosity function.

There are several algorithms to measure the luminosity function
(see \citet{Will97} for a review). The non-parametric methods (e.g.
\citealt{LB71}) usually require certain uniformity of data coverage
to be applicable. The likelihood-based methods \citep{MTA83} or
Bayesian methods \citep{KFV08, CLM13} are preferable for the FRB
problems, because these algorithms are more flexible in modelling
the systematics and less constrained by the conditions of a given
sample.

In this paper, we aim to measure the normalised FRB luminosity
function. To include  systematics and unknowns in the statistical
inference, we develop a Bayesian framework suitable for the current
problem. For most of the known FRBs, there are four main
observables relevant to the luminosity function determination: flux density, 
bandwidth,
duration, and dispersion measure. Compared to the other astronomical sources
whose luminosity functions are measured, the FRB distance is not
directly available. As a result, we have to rely on the dispersion measure
to indirectly infer the FRB distance. Our method to measure the FRB luminosity 
function includes three major steps: (1) mitigate the Galactic
foreground contribution of the dispersion measure; (2) model the FRB
host galaxy and the cosmological dispersion measure contribution; (3) include 
dispersion measure models in the Bayesian luminosity function inference, and 
marginalise the unknowns. The first step is straightforward,
as good knowledge on the Galactic electron distribution is available.
The second step is to model the effects of some unknown properties on determining the 
luminosity function. The third step is to use a Bayesian method developed in this paper to 
`enumerate' all possibilities and include the unknowns in the statistical 
inference. We can then determine the contribution of the
unknowns to statistical errors, e.g. we can make sure that the confidence 
bounds of inferred parameters contain the uncertainties in the modelling.

The paper is organised as follows.  In \SEC{sec:frbobs}, we explain
how we remove the dispersion
measure contribution from the Galactic foreground.  In \SEC{sec:meth}, we 
describe our Bayesian
inference method.  The likelihood function is built in \SEC{sec:likf},
with detailed modelling of its components in the rest of the subsections of 
\SEC{sec:meth}. The computational method for
posterior evaluation is shown in \SEC{sec:post}. Our results are
given in \SEC{sec:res}, with discussion made in \SEC{sec:disc}. For
the readers' convenience, we summarise the symbols used
throughout this paper in \TAB{tab:not}.

\section{Pre-processing the FRB data}
\label{sec:frbobs}

For most FRBs, the measured parameters are peak flux density ($S_{\rm 
peak}$), burst duration ($w$), and dispersion measure \begin{equation}
	\DM=\int n_{\rm e} \D l \,,
\end{equation}
i.e., the electron density $n_{\rm e}$ integrated along the line of sight, which 
serves as the distance indicator for the FRBs.

When radio waves propagate through interstellar medium (ISM), the group velocity 
becomes frequency-dependent \citep{LandauEM}. For the rest-frame observer, the 
time
delay between the pulses at two different frequencies is
\begin{equation}
\Delta t =4.15\, {\rm ms}\, \left(\frac{\DM}{1\, \cmpc}\right)
	\left[\left(\frac{\nu_1}{1\, \mathrm{GHz}}\right)^{-2} -\left(\frac{\nu_2}{1\, 
	\mathrm{GHz}}\right)^{-2}\right] \,,
	\label{eq:delaydef}
\end{equation}
under the assumption that the radio wave frequency is higher than the ISM plasma 
frequency.  The DMs are then usually measured by fitting the observed time 
delays using \EQ{eq:delaydef}.

All the data used in this paper comes from the FRB
catalogue (FRBCAT)\footnote{http://frbcat.org/} compiled by \citet{Petroff16PASA}
amended with the original discovery papers. In \TAB{tab:frbs} of 
\APP{app:dattab},
We list the values of the observed and inferred parameters of FRBs used in the 
current paper for reader's reference.

The DM of an FRB has contributions from five components, i.e.
\begin{equation}
\DM = \DMm+\DMd+\DMi(z)+\frac{\DMh+\DMs}{1+z}.
\end{equation}
In the above expression
 $\DMm$ is the component due to the Milky Way free electrons,  $\DMd$ is
the possible component contributed by the electron halo of the Milky Way,
$\DMi$ is the intergalactic medium (IGM) contribution,   
$\DMh$ is the FRB host galaxy contribution, and $\DMs$ is the component from the local environment 
surrounding the FRB source in small scales, e.g. H\textsc{ii} regions, ionised gas halos, 
magnetospheres. The cosmological redshift factor, $1+z$, converts the DM seen by 
the rest-frame observer to that of the Earth observer as shown by \cite{DZ14ApJ}.

There are currently two models that are widely used for the
Galactic distribution of free electrons:  NE2001 \citep{CL02}, and
YMW16 \citep{YMW16}.  The NE2001 model contains several components
for the electron density distribution, the thin and thick asymmetric
disks, the spiral arms, a local arm, a local hot bubble surrounding
the Sun, and relatively large super-bubbles
 in the first and third Galactic quadrants. It also includes over
 dense components representing the small scale structures.  By
 contrast, the more recent YMW16 model contains a
four-armed spiral pattern together with the local structures similar
to that of NE2001.  YMW16 does not include the clumps or voids to
correct for DMs of individual pulsars, but more pulsars with
independent distance measurements were used in fitting the model
parameters.  Compared with that of NE2001, the average electron
density of YMW16 is lower \citep{YMW16}.

In our data preprocessing, we remove the Milky Way contribution from the
observed DM of each FRB to get the extragalactic contribution based on
two representative models described above (i.e. the NE2001 and YMW16).
The observed DM as well as the extragalactic DM ($\DMe$), i.e. 
$\DMe=\DM-\DMm$, are listed in \TAB{tab:frbs}. As one
can see, most of the extragalactic DM values are compatible between
the two Galactic electron models; Only for certain FRBs, e.g. FRB
010621, there is a factor-of-two difference.  

The Milky Way dark halo
may contribute to the DM. The standard picture \citep{SWS03,BL07,
GM08}, however, indicates a very low electron density ($n_{\rm e}<
10^{-3}-10^{-4}\, \rm cm^{-3}$) in the extended Milky Way halo with
typical predictions of $\DMd\simeq30\, \cmpc$ \citep{Dolag15MN}.  We
compare the results with and without correcting the halo contribution in 
\SEC{sec:res}, i.e. the results using $\DMe=\DM-\DMm-\DMd$ and $\DMe=\DM-\DMm$.
The negligible difference in the results legitimate performing the halo 
correction in the pre-processing stage {\it a posteriori} and save us from the 
complex probabilistic modelling. However we are not that lucky for other 
systematics, which requires proper modelling as shown in the next section. 

\section{Bayesian framework to measure the FRB luminosity function}
\label{sec:meth}
We develop a Bayesian data analysis scheme to measure the 
luminosity function of known FRBs from three observables, the peak flux density, 
the burst duration and the extragalactic DM.
These observables are insufficient to directly compute the FRB luminosity, 
because the FRB distance and DM do not fall into the one-to-one relation.  In 
order to measure the luminosity function, we seek help from the Bayesian method, 
which can include the systematics of the unknowns.
Bayesian inference \citep[see, e.g.,][for details]{Jaynes03} helps to
convert the `probability of data' to the `probability of parameters' via Bayes' 
theorem, \begin{equation} P({\bf
\Theta|X})=\frac{P({\bf \Theta})P({\bf X|\Theta})}{P({\bf X})}\,,
\label{eq:bayest}
\end{equation}
where $\bf X$ represents the \emph{data}, and $\bf \Theta$ is a vector set of model \emph{parameters} to 
be inferred. The \emph{likelihood} function, $\Lambda \equiv P({\bf X|\Theta})$,  
is the PDF of the data given the model parameters.  $P(\bf \Theta|X)$ is the 
\emph{posterior} PDF, i.e. the PDF for the parameters given the data set.  The 
Bayesian \emph{evidence} $P(\bf X)$ is a normalization coefficient that
\begin{equation}
	P(\bf X)=\int P({\bf \Theta})P({\bf X|\Theta}) d\VEC{\Theta}\,.
	\label{eq:evid}
\end{equation}
The \emph{prior} PDF $P(\bf \Theta)$ describes our information {\it a priori} 
about the model parameters. In the current paper, the data $\VEC{X}$ are the 
measured FRB parameters (i.e. ${\rm DM}_{\rm E}$, $S_{\rm peak}$, and $w$), and 
the parameters $\VEC{\Theta}$ are for the luminosity function.  In the common 
practice of Bayesian data analysis,
one needs to construct the likelihood function and compute the posterior to 
infer the parameters.

\subsection{Likelihood function}
\label{sec:likf}
We construct the likelihood function under six assumptions.

{\bf i)} The FRB luminosity distribution follows the Schechter function 
\citep{Schechter76ApJ}, which was widely used for galaxies, quasars and 
gamma-ray bursts. It takes the form of
\begin{equation}
\phi(\log L)\D \log L=\phi^* 
\left(\frac{L}{L^*}\right)^{\alpha+1}e^{-\frac{L}{L^*}}\,\D \log L\, ,
\label{eq:sf}
\end{equation}
where $\phi^*$ is the normalisation factor, $\alpha$ is the power-law index of 
the distribution and $L^*$ is the cut-off luminosity. There are two considerations 
to use the Schechter function. Firstly, the function includes a common power-law  
function with the inclusion of an exponential cut-off. In the r\'egime $L<L^{*}$,
the function is consistent with a power law. The cutoff ensures that there exists
a maximal luminosity of FRBs. Second, such a function has been used in describing
the luminosity functions of other astrophysical objects.

{\bf ii)} The cosmological evolution of FRB luminosity function can be 
neglected, in other words, the parameters in the Schechter function are 
independent of redshift.

{\bf iii)} The spatial distribution of FRBs is homogeneous in the comoving volume,
i.e. the PDF for the comoving radius $r$ proportional to the differential 
comoving volume, i.e. $f_{\rm
r}(r)\propto dV/dr\propto r^2$. As a caveat, it is well known \citep{BST88} that 
the source may not be perfectly homogeneous in the comoving volume.  
Particularly, one needs to factor in the effects of luminosity function and 
redshift distribution (see Equation (15) in \citet{BST88}).
However, we had only the limited number of FRBs, the homogeneous assumption is a valid `first-order' 
approximation.  Tests for homogeneity are only possible when a sufficient 
number of FRBs are detected.

{\bf iv)} The luminosity distribution of FRBs is independent of FRB positions in 
their host galaxies. 

{\bf v)} The source DM contribution ($\DMs$) is independent of the host galaxy 
dispersion measure and the FRB luminosity, i.e. $\DMs$ is independent of $\DMh$ 
and $L$. Here, the $\DMs$ is dedicated to the local environment of FRBs, of 
which the sizes are much smaller than the host galaxy. The host-galaxy-dependant DM
in our modelling is through $\DMh$ as discussed in \SEC{sec:dms}.

{\bf vi)} The FRB true position distributes uniformly (per solid angle) inside the telescope main beam. 
The off-centre position introduces a lower beam response with
$\epsilon\le1$ (See \SEC{sec:beam})

With the above six assumptions, FRB luminosity ($L$), comoving radius ($r$), 
host galaxy DM, FRB local DM ($\DMs$), and beam response ($\epsilon$)
become independent random variables. Thus the joint PDF becomes multiplicative, 
i.e
\begin{equation}
\begin{split}
	f(\log L, r, \DMh, \DMs, \log\epsilon)&=\phi(\log L)\, \fr(r)\,\fd(\DMh|z) \\
	&\quad \times  \fs(\DMs)\, \fe(\log\epsilon)\,
\end{split}
\label{eq:lik1}
\end{equation}
where $\fs$ is the PDF of $\DMs$, and $\fe$ is the PDF of beam response of radio 
telescope.
The free electron density in the host galaxies highly depends on the star 
formation activity, which is roughly reduced by a factor of 10 from redshift $z=1$ to 
$z=0$ \citep{hb06apj, MD14}.
The PDF $\fd(\DMh|z)$ for the rest-frame $\DMh$ becomes redshift dependent.

To compute the likelihood, we need to obtain the PDF of the observables. This 
can be done by the nonsingular random variables transformation \citep{Fisz63}.
We map the PDF of quintet $\{\log L, r, \DMh, \DMs, \log\epsilon \}$ to that of $\{\log 
S,\DMe, z, \DMs, \log\epsilon \}$ using the Jacobian transformation. As a
nonsingular transformation, one has \begin{equation}
\begin{aligned}
	f(\log S,\DMe, z, \DMs, \log\epsilon)&=\left| \frac{\partial (\log L, r, \DMh, \DMs, \log\epsilon)}{\partial (\log 
	S,\DMe, z, \DMs, \log\epsilon)}\right| \\
	&\quad \times f(\log L, r, \DMh, \DMs, \log\epsilon)\,.
	\label{eq:jdet}
\end{aligned}
\end{equation}
The Jacobian determinant is calculated using the luminosity-flux and 
$\DMe$-($\DMh$,$\DMs$) relations as follows.  

The apparent flux density is determined by source luminosity and beam response, i.e.
\begin{eqnarray}
	\log S&=&\log L-2\log r_{\rm L} -\log {\rm \Delta \nu_0} +\log \epsilon -\log 
	4\pi\,, \label{eq:l2s} \\
	r_{\rm L}&=&(1+z)r\,,  \label{eq:z2r}\\
	r&=&\frac{c}{H_0} \int_{0}^{z} \frac{1}{E(z)}\, \D z\,. 
\end{eqnarray}
Here we assume the intrinsic spectrum of FRB is flat, and the spectral width 
$\Delta \nu_0$ is fixed to the reference values of 1 GHz. The Hubble constant 
is taken as $H_{0}=67.8 \, {\rm km\,s^{-1}\,Mpc^{-1}}$ \citep{Planck16A&A}.  The luminosity 
distance, $r_{\rm L}$, is computed from the comoving distance, $r$.  The 
function
\begin{equation}
	E(z)=\sqrt{\Omega_{\rm m}(1+z)^3+\Omega_{\Lambda}}
	\label{eq:funce}
\end{equation}
is the logarithmic 
time derivative of the cosmic scale factor in a flat $\Lambda$CDM universe ($\Omega_k\simeq0$), in which we adopt
dimensionless matter density $\Omega_{\rm m}=0.308$ and cosmological 
constant $\Omega_{\Lambda}=0.692$ \citep{Planck16A&A}.

The intrinsic DM from the host galaxy is calculated by subtracting the IGM 
and source contributions from the extragalactic DM, i.e.
\begin{equation}
	\DMh=(\DMe-\DMi) (1+z)-\DMs \,, \label{eq:dme2dmh}
\end{equation}
where the factor $(1+z)$ comes from converting the DM seen by the Earth observer
to the DM seen by the FRB rest-frame observer \citep{Ioka03ApJ, Inoue04MN}.
The electron density of IGM depends on the ionization history of the Universe 
(\citealt{DZ14ApJ}, see also \APP{sec:dmi})
\begin{equation}
		\DMi \simeq 1.1\times 10^{3} \int_0^z\frac{f_{\mathrm{IGM}}\, g(z)\, (1+z)\D z}{E(z)}\, 
		\cmpc\,,
\label{eq:dmi}
\end{equation}
where the $f_{\mathrm{IGM}}$ is the cosmological baryon mass fraction in the IGM, 
here we adopt $f_{\mathrm{IGM}}\simeq0.83$ from the summation of global budget 
of baryons in all states \citep{Fukugita98ApJ}. The function $g(z)$, on the 
right hand of \EQ{eq:dmi}, is the ionised electron number fraction per baryon.
One has
\begin{equation}
g(z)\simeq\frac{3}{4}\chi_{e,\mathrm{H}}(z)+\frac{1}{8}\chi_{e,\mathrm{He}}(z)\,,
\end{equation}
where $\chi_{e,\mathrm{H}}$ and $\chi_{e,\mathrm{He}}$ are the cosmic ionisation 
fraction of hydrogen and helium, respectively. FRBs are located relatively nearby, so that one 
can safely adopt $\chi_{e,\mathrm{H}}\simeq1$ and $\chi_{e,\mathrm{He}}\simeq1$
\citep{FCK06AR, McQuinn09ApJ}.

Using \EQ{eq:l2s}, (\ref{eq:z2r}) and (\ref{eq:dme2dmh}), we calculate the 
Jacobian determinant in \EQ{eq:jdet}. After maginalisation of $\DMs$ and 
$\epsilon$, the PDF $f(\log S,\DMe, z)$ becomes (see \APP{app:bayes} for 
details).  \begin{equation}
				 f(\log S, \DMe, z) = I(\log L)\, \fz(z)\, I(\DMe, z)\, (1+z)\, ,
 \\
\end{equation}
where the marginalisations for the unknown source DM ($\DMs$) and beam response 
($\epsilon$) are \begin{equation}
\begin{aligned}
I(\DMe, z) &\equiv \int_0^{\max(\DMs)}\fd(\DMh|z) \fs(\DMs)\, \D\DMs \, ,
\end{aligned}
\end{equation}
and
\begin{equation}
\begin{aligned}
	I(\log L)\equiv\int \phi(\log L) \fe(\log\epsilon) \D\log\epsilon
\end{aligned}
\end{equation}

Since only one FRB has a measured redshift so far, we need to marginalise the 
redshift in the likelihood to include such an ignorance. The reduced likelihood 
function, as what will be used in the Bayesian inference, is
\begin{equation}
\begin{aligned}
	f(\log S,\DMe)&=\frac{1}{N_{\rm f}}\int_0^{\infty}
	I(\log L)\, \fz(z)\, I(\DMe, z) \, (1+z)\,\D z ,
\end{aligned}
	\label{eq:likfun}
\end{equation}
where $\fz(z)$ is the FRB spatial distribution function in the redshift 
space with
\begin{equation}
	\fz(z)=\frac{r(z)^2}{E(z)}\,,
	\label{eq:fz}
\end{equation}
and $N_f$ is the normalisation factor as
\begin{equation} N_f = \int_{\log S_{\rm min}}^{\infty}\,\D \log S \int \int f(\log 
	S, \DMe, z)\, \D \DMe\, \D z.  \\ \label{eq:norm}
	\end{equation}
The lower limit of the flux density integration, $S_{\rm min}$, is the minimum 
detectable flux density of the telescope at the time when the given FRB was 
detected, i.e.~the survey depth.  The radiometer equation \citep{LK12HPA} gives
\begin{equation}
	S_{\rm min}=\frac{\mathrm{S/N}_0\ T_{\mathrm{sys}}}{G\sqrt{N_\mathrm{p}\,\mathrm{BW}\, w}}=\frac{\mathrm{S/N}_0\, {\rm SEFD}}{\sqrt{N_\mathrm{p}\, \mathrm{BW}\, w}},
	\label{eq:rdm}
\end{equation}
where $w$ is the FRB pulse width, $\mathrm{S/N}_0$ is the signal-to-noise ratio threshold 
for detection in the surveys, $T_{\rm sys}$ is the system temperature, 
$G$ is the telescope gain, $N_\mathrm{p}$ is the number of polarisations summed, and 
${\rm BW}$ is the bandwidth. The system temperature and gain can be combined 
using the system equivalent flux density (${\rm SEFD}\equiv T_{\rm sys}/G$) as shown on the right-hand side 
of \EQ{eq:rdm}. The parameters for the depths of surveys are given in 
\TAB{tab:svy}, the numeric values of the corresponding parameters are from the 
reference listed in \TAB{tab:frbs}.

\begin{table}
	\caption{The instrumental parameters of FRB surveys}
\centering
\begin{threeparttable}
\begin{tabular}{cccccccc}
\hline
\hline
Survey & $G$ & $T_{\rm sys}$\tnote{a} & $\rm SEFD$& BW & S/N$_{0}$ & $N_\mathrm{p}$ & Ref.\tnote{b}\\
 &K/Jy & K & Jy & MHz & & &\\
\hline
Parkes I & 0.69 & 28 &41& 288 & 7 & 2 & [1] \\
Parkes II & 0.69 & 28 &41& 338 & 10 & 2 & [2]\\
Arecibo & 0.7\tnote{c} & 30 &43& 322 & 7 & 2 & [3]\\
GBT & 2.0 & 25 &13& 200 & 8 & 2 & [4] \\
UTMOST & 3.0 & 400 &130& 16 & 10 & 1 & [5]\\
ASKAP &n/a &n/a &1800& 336 & 10 & 2 & [6]\\
\hline
\end{tabular}
\begin{tablenotes}
\footnotesize
\item (a) For different FRB detections, the telescope system temperatures depend on the detected beams. Hence, in the calculation for sensitivity of each FRB, we adopted the corresponding value from the newest FRBCAT.
\item (b) The references are: [1] \cite{Lorimer07Sci}; [2] \cite{Thornton13Sci}; [3] \cite{Spitler14ApJ}; [4] \cite{Masui15Nat}; [5] \cite{Caleb17MN}; [6] \cite{Bannister17ApJ}.
\item (c) The Arecibo FRB was detected probably in the sidelobe of multibeam receiver, the gain of 
sidelobe is taken as 0.7 K/Jy \citep{Spitler14ApJ}. 
\end{tablenotes}
\label{tab:svy}
\end{threeparttable}
\end{table}

We need to model the beam response $\fe(\log\epsilon)$, local DM and host galaxy 
DM distribution function $\fs(\DMs)$ and $f_{\cal D}(\DMh|z)$ before computing 
the likelihood. The modelling will be explained in the next sections.

\subsection{The beam response likelihood}
\label{sec:beam}
We can approximate the main-beam response using a Gaussian function 
\citep{BWOptics}, where the ratio  between the
observed flux and the intrinsic flux of an FRB,
\begin{equation}
	\epsilon\equiv\frac{S_{\rm obs}}{S_{\rm src}}=e^{-4 \ln 2 
	\left(\frac{\theta}{\theta_{\rm b}
	}\right)^2}\,.
\end{equation}
In this expression, $S_{\rm src}$ and $S_{\rm obs}$ are the true and observed flux of FRB.
$\theta$ is the angular distance between
the true position of FRB and the beam centre. $\theta_{\rm b}$ is the 
full-width-half-maximum (FWHM) beam size, i.e. $\epsilon=0.5$ for
$\theta=\theta_{\rm b}/2$.

If we assume a uniform PDF per solid angle for the source position inside the 
main beam, i.e.
accepting Assumption vi) made in \SEC{sec:likf}, the PDF
of $\cos\theta$ will also be uniform. For most of the radio telescopes, if not 
all, $\theta_{\rm b} \ll 1\,\mathrm{rad}$, so does $\theta$, including the telescopes that have beams with large semi-major axis and small semi-minor axis, e.g. UTMOST. Thus $\theta^2$ also follows 
a uniform PDF. As $\log \epsilon \propto -\theta^2$, the PDF of $\log 
\epsilon$ is uniform as well. We get
$\fe(\log\epsilon)={\rm constant}$.

\subsection{The distribution function for the local DM of the FRB source}
\label{sec:dislocdm}

The nature of FRB origins is still under debate and the PDF of FRB local DM  is 
highly uncertain. Investigations \citep{Yu14ApJ, Cao17ApJ} had shown that the DM 
contribution from a pulsar wind is less than 10 $\cmpc$ for a reasonable range of 
pair multiplicity parameter. The DM of such origins can be even smaller, as the 
electrons close to the FRB should be relativistic and contribute little to the 
DM \citep{LP82PhST, GBI06PPM}.  However, optical observations have shown that the
repeating FRB 121102 is in a star-forming region \citep{Kokubo17ApJ, Bassa17ApJ} and the 
source DM may not be negligible \citep{Yang17ApJ}. In this paper, we take a 
least-informative assumption \citep{Jaynes03} that $\DMs$ follows a uniform PDF 
in a rather wider range from 0 to 50 $\cmpc$. In this way, we incorporate the 
unknowns into the error of inferred parameters.

\subsection{The PDF of the FRB host galaxy DM}
\label{sec:dms}

\citet{XH15RAA} have modelled the FRB host DM distribution
assuming that the host galaxies are Milky Way-like or M31-like. In
our work, we use Monte Carlo simulations to calculate the rest-frame
DM distribution function, i.e. the DM distribution function as seen
by observers local to the galaxies. Compared with \citet{XH15RAA},
instead of focusing on specific galaxies, we study the galaxy
population and build an ensemble DM PDF.  That
is, we want to determine how the DM distribution of FRBs looks like
for a galaxy-rest-frame observer.  The summary for our recipe is
as follows:

{\bf i)} For a given galaxy type, we simulate the $\Ha$ and $r$-band luminosity 
for one galaxy each time according to the galaxy $\Ha$ and $r$-band luminosity 
function \citep{Nakamura03AJ, Nakamura04AJ} at the zero redshift. The details 
are described in \SEC{sec:spg}.

{\bf ii)} Using the simulated values of $\Ha$ and $r$-band luminosity from the 
step i), we simulate one DM value for an FRB in the galaxy for a rest-frame 
observer. Here, the DM value is computed by scaling from the `template 
galaxies', where the DM distribution of template galaxies
are calculated in \SEC{sec:galtemp}.
The scaling between the $\DMh$ of two galaxies of the same type for the given 
line of sight depends on the size of the galaxy and electron density, where (see 
\SEC{sec:spg}),
\begin{equation}
	\frac{ {\rm DM}_{\rm host, 1}}{ {\rm DM}_{\rm 
	host,2}}\propto\sqrt{\frac{L_{\rm \Ha, 1} R_\mathrm{e, 2}}{ L_{\rm \Ha,2} 
	R_\mathrm{e, 1}}}\,.
\end{equation}
Here, $L_{\rm \Ha}$ is the $\Ha$ luminosity, $R_\mathrm{e}$ is the effective radius of 
galaxy being derived from the $r$-band luminosity. 

{\bf iii)} We repeat the steps `i' and `ii' for one million times and use the 
accumulated DM values to build the DM distribution function. The analytic form 
of the distribution function is then derived by curve fitting. The DM 
distribution function ($f_{\cal D}(\DMh)$) at this stage is the 
rest-frame-zero-redshift distribution function, because we compute the DM value 
for the rest-frame observers using the nearby galaxy luminosity function. 

{\bf iv)} We convert the \emph{rest-frame-zero-redshift} DM distribution 
function to the \emph{rest-frame} DM distribution function to accommodate the 
evolution of star formation history. As the $\Ha$ luminosity scales with the 
star formation rate ($\rm SFR$, see \citealt{KTC94ApJ, MPD98ApJ} ), the electron 
density $n_{\rm e}$ becomes ${\rm SFR}$-dependent that $n_{\rm e}\propto {\rm 
SFR}^{1/2}$ (see \APP{app:scaling}).  The  rest-frame DM distribution 
function at redshift $z$ then becomes \begin{equation}
	\fd(\DMh|z)= \sqrt{\frac{\mathrm{SFR}(0)}{\mathrm
{SFR}(z)}}\fd\left[\DMh \sqrt{\frac{\mathrm{SFR}(0)}{
	\mathrm{SFR}(z)}}\right]\,,
\end{equation} i.e. $\fd(\DMh|z)$ is the distribution function of 
$\DMh$ at the redshift $z$ measured by the rest-frame observers also at the 
redshift $z$.  Here the function $\fd[\cdot]$ on the right-hand side of 
the equation is the zero-redshift-rest-frame DM distribution function from the 
step iii). The star formation history we used \citep{hb06apj} is 
\begin{equation}
	{\rm SFR}(z)=\frac{0.017+0.13 z}{1+(z/3.3)^{5.3}}\, {M_{\sun}\, \rm yr^{-1}}\, \rm Mpc^{-3}\,.  \label{eq:sfrz}
\end{equation}

\subsubsection{DM scaling via host galaxy $\Ha$ and $r$-band luminosity}
\label{sec:spg}

The average electron density is computed from $\Ha$ luminosity 
(\APP{app:scaling}) with
\begin{equation}
\begin{split}
	\left\langle n_{\rm e} \right\rangle&=1.0\, \eta^{2/3}\left(\frac{T}{10^4\ 
	\mathrm{K}}\right)^{0.45}\left(\frac{L_{\Ha}}{10^{40}\,\ergs}\right)^{1/2} \\
	& \quad \left(\frac{R}{1\, \kpc}\right)^{-3/2}\, \mathrm{cm}^{-3},
\end{split}
\label{eq:nea}
\end{equation}
where $\eta$ is the filling factor, $T$ is the ionised gas temperature, and $R$ 
is the galaxy radius. The typical electron temperatures in galaxies are in a 
rather narrow range from 5,000 K to 10,000 K.  Due to the flat $0.45$ index, we 
fix the gas temperature to 8,000 K, which leads to at most 20\% error in 
determining $n_{\rm e}$.

Because $\DMh\propto n_{\rm e} R_{\rm e}$, the scaling relation between the 
$\DMh$ values of two galaxies for the line of sight along the same directions becomes
\begin{equation}
	\frac{ {\rm DM}_{\rm host, 1}}{ {\rm DM}_{\rm host,2}}=\frac{\langle n_{\rm 
	e}\rangle_1  R_\mathrm{e,1}}{ \langle n_{\rm e}\rangle_2  
	R_\mathrm{e, 2}}=\sqrt{\frac{L_{\rm \Ha, 1} R_\mathrm{e, 2}}{ L_{\rm \Ha,2} 
	R_\mathrm{e, 1}}}\,.
	\label{eq:scaldm}
\end{equation}
In this way, once we know the DM value of a template galaxy, we can calculate 
the DM value of
another galaxy of the same type by using the above scaling equation. The unknown 
filling factor in \EQ{eq:nea} is canceled, assuming
it is a constant for all the galaxies with the same type. The template galaxy is not 
necessarily a typical member of the given type and merely serves as a 
reference. We delay the discussions on the template galaxy to the next section, 
and focus on the distribution functions of $L_{\rm \Ha}$ and $R_{\rm e}$ at the 
moment.

\begin{figure}
\centering
\includegraphics[width=3.5in]{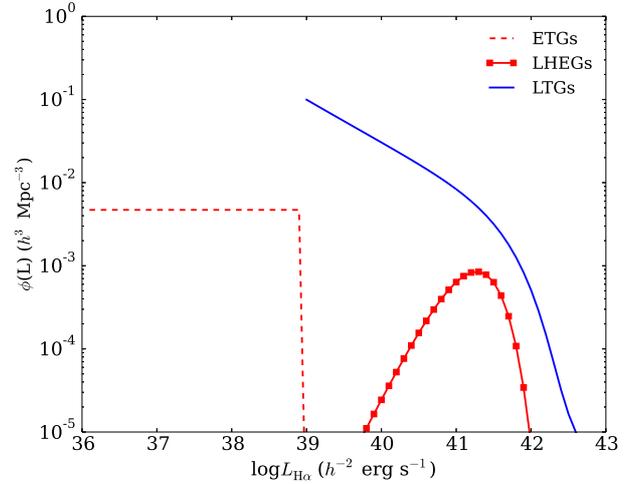}
\caption{The $\Ha$ luminosity functions for ETGs and LTGs. The solid curve 
(blue) is for the LTG luminosity function. For the luminosity of ETGs, we 
extended the original results of \citet{Nakamura04AJ} (in red 
curve with square marks), where our extension is plotted as the red dashed curve.  
The details of extension operation are described in the main text.}
\label{fig:LFs}
\end{figure}

$L_{\rm \Ha}$ is simulated according to the $\Ha$ luminosity functions. Based on 
the complete survey data from the \emph{Sloan Digital Sky Survey} with a 
redshift depth of $z=0.12$, \citet{Nakamura04AJ} measured the morphologically 
classified $\Ha$ luminosity functions. The $\Ha$
luminosity functions for the early-type galaxies (ETGs, with morphological index 
$T_\mathrm{morph}\le1.0$ as defined by \citealt{Nakamura04AJ}) and the 
later-type galaxies (LTGs, with morphological index $T_\mathrm{morph}>1.5$) take 
forms of
\begin{eqnarray}
	\phi_{\rm ETG}(L_{\Ha})&\propto& 0.8\left(\frac{L_{\Ha}}{10^{40.02}}\right)^{0.79} 
	e^{-
	\frac{L_{\Ha}}{10^{40.02}}}\,,  \label{eq:haltg}\\
	\label{eq:haetg} \phi_{\rm
	LTG}(L_{\Ha}) &\propto&1.0\left(\frac{L_{\Ha}}{10^{41.7}}\right)^{-1.4} e^{-
	\frac{L_{\Ha}}{10^{41.7}}} \nonumber \\ &+&1.0
	\left(\frac{L_{\Ha}}{10^{41.7}}\right)^{-1.53} e^{-
	\frac{L_{\Ha}}{10^{41.71}}} \nonumber \\ &+&0.01
	\left(\frac{L_{\Ha}}{10^{42.8}}\right)^{-1.77} e^{-
	\frac{L_{\Ha}}{10^{42.8}}} \,.  \end{eqnarray}
Here we summed the luminosity functions of the sub-types to form the luminosity 
functions of LTGs. The functions are plotted in \FIG{fig:LFs}.

As shown in \FIG{fig:LFs}, the ETG $\Ha$ luminosity function of
\citet{Nakamura04AJ} peaks around $L=10^{41}\, \ergs$.  As a common
cherished belief (e.g.  \citealt{Kenn98}), most of the ETGs are the
old galaxies with little star forming activity and hence with low $\Ha$
luminosities. However, the results of \citet{Nakamura04AJ} indicate
that the average $\Ha$ luminosity of ETGs would be higher than that
of LTGs. Such discrepancy is mainly due to selection effects, that
the low $\Ha$ luminosity galaxies were invisible in the survey and
the \emph{luminous $\Ha$ elliptical galaxy} (LHEG) contributions
start to bias the results.  Indeed, \cite{Nakamura04AJ} mentioned
if the sample selection criterion they used is strong enough, they
would reject 235 AGNs, which is 35\% in the current $\Ha$ detected
sample including both the ETGs and the LTGs. To compute the missing
fraction of ETGs in the $\Ha$ selected sample, we compare the $\Ha$
luminosity functions with the \textit{r}-band luminosity functions
\citep{Nakamura03AJ} of the same sample. The integrated ETG volume density
using the \textit{r}-band luminosity function is $n^*=2\times10^{-2}\,
h^3\, {\rm Mpc}^{-3}$, while the galaxy volume density produced by the
$\Ha$ luminosity function is only $n^*=7.2\times 10^{-4}\, h^3\, {\rm
Mpc}^{-3}$.  Clearly, the majority (more than 90\%) of ETGs are
below the detection limit in the $\Ha$ selection. We thus regard the original
$\Ha$ luminosity function of ETGs in \citet{Nakamura04AJ} only applicable for 
LHEGs.

In order to get the $\Ha$ luminosity distribution for the full ETG
population, an extension operation is needed. As little information
is available for these low $\Ha$ luminosity galaxies, we perform
the most na\"ive correction. We assume that the missing ETGs
distribute uniformly in logarithmic luminosity space below the
survey sensitivity. Such an extension scheme introduces the least
amount of information as being widely applied in the Bayesian
statistics \citep{Jaynes03}. The extension leads to a constant
density $\phi=4.7\times 10^{-3}\, h^3\,{\rm Mpc^{-3}}$ in the range
of $10^{36}\,\ergs<L<10^{39}\,\ergs$.

We now turn to the distribution of galaxy radii. The galaxy radius can be 
calculated from the optical luminosity using empirical size-magnitude relations
\citep{Shen03MN}, that
\begin{equation}
	\log{\left(\frac{\bar R_{50}}{\kpc}\right)} = \begin{cases}
	-0.4aM+b\,, \quad\quad\quad\quad \textrm{for ETGs} \\
-0.4\alpha M+(\beta-\alpha)
\log{[1+10^{-0.4(M-M_0)}]} \\
+\gamma,  \quad\quad \quad\quad \quad\quad\quad\quad\textrm{for LTGs}\,.
\end{cases}
\label{eq:size}
\end{equation}
We sample the $r$-band optical magnitude ($M$) based on the luminosity functions 
of \citet{Nakamura03AJ} and use the above size-magnitude relations to compute 
the Petrosian half-light radius ($R_{50}$, see \citealt{Petrosian76ApJ}), where 
constants $a=0.60$, $b=-4.63$ for ETGs and $\alpha=0.21$, $\beta=0.53$, 
$\gamma=-1.31$, $M_0=-20.52$ for LTGs. We then convert the Petrosian radius to 
the effective radius \citep{Graham2005AJ....130.1535G}, that, for ETGs
$R_{50}=0.73\, R_\mathrm{e}$, and, for LTGs $R_\mathrm{50}=0.99\, R_\mathrm{e}$.

To confirm the validity of above modelling, we compare the estimated
values with the observations. For the LTG, the measured
stellar-density-weighted electron density of the Milky Way by the YMW16
 is $\langle n_{\rm e} \rangle=0.04\,{\rm cm^{-3}}$, while the current modelling 
 ($R_{\rm e}\simeq 3.5\, {\rm kpc}, L_{\rm
\Ha}\simeq5\times10^{40}\, {\rm erg\,s^{-1}}$) produces $0.014\sim0.066\,
{\rm cm^{-3}}$ when we adopt the filling factor from 0.01 to 0.1.  
For the ETG type, the measured average free electron density of M87 from 
Chandra observations  \citep{Cavagnolo09ApJS} is 
$\langle n_{\rm e}\rangle=0.05\, {\rm cm^{-3}}$, and the modelled electron 
density ($R_{\rm e}\simeq 7.7\, {\rm kpc},
L_{\rm \Ha}=10^{40}\,\ergs$) is $0.009\sim0.042\,{\rm cm^{-3}}$.
Clearly, the predictions for the electron density depend on the filling factor. However, 
since we are using the scaling relation, \EQ{eq:scaldm}, to compute the DM of the
simulated galaxies, the filling factors cancel out. In this case, the results will not be 
affected.

As a short summary for this section, we create a large sample of artificial 
galaxies, in which $L_{\Ha}$ follows the morphological luminosity function 
(\EQ{eq:haltg} or (\ref{eq:haetg})) and radius follows the size distribution in 
\EQ{eq:size}. We then convert the $\DMh$ of a template galaxy (see below in \ref{sec:galtemp}) 
to that of the given galaxy according to \EQ{eq:scaldm}.  The DM distribution of the template
galaxies and galaxy ensembles will be discussed in the next section.

\subsubsection{DM for the template galaxies} \label{sec:galtemp}

In this section, we compute the DM distribution of the template galaxies, where 
the stellar distribution and electron density modelling of galaxies are
considered. Due to the morphological difference, we need to address the ETGs, 
LHEGs, and the LTGs separately.

\emph{ETGs} and \emph{LHEGs}: The electron density model of ETGs, 
unfortunately, is not well studied statistically, particularly due to 
the low gas fraction.  Also one usually needs galaxies with larger angular 
diameters,  that can be resolved in order to measure the electron 
distribution. As a result, there will be only a few ETGs with electron density 
profile measurements, and those ETGs might not fall into the class of 
stereotype. However, as explained above, 
since our DM scaling 
relation accounts for both  galaxy size and luminosity, we can use any 
individual member as the reference.  As a caveat, we need to assume that the gas 
filling factor varies only mildly
in the galaxy population. In the paper, M87 is chosen as the reference, 
simply because it has a published electron density profile 
\citep{Cavagnolo09ApJS}.

The electron density profile of M87
derived from Chandra surface brightness measurements \citep{Cavagnolo09ApJS}
can be characterised by the following function 
\citep{FG83ApJ}
\begin{equation}
	n_e=n_0\left[1+\left ( \frac{R}{R_0}\right)^2\right]^{\alpha_\mathrm{e}}\,,
	\label{eq:ne_etg}
\end{equation}
where the fitted parameters are $n_0\simeq0.165\ \mathrm{cm}^{-3}$, $R_0=1.544\,
{\rm kpc}$, and $\alpha_\mathrm{e}=-0.582$.

\begin{figure} \centering \includegraphics[width=3.5in]{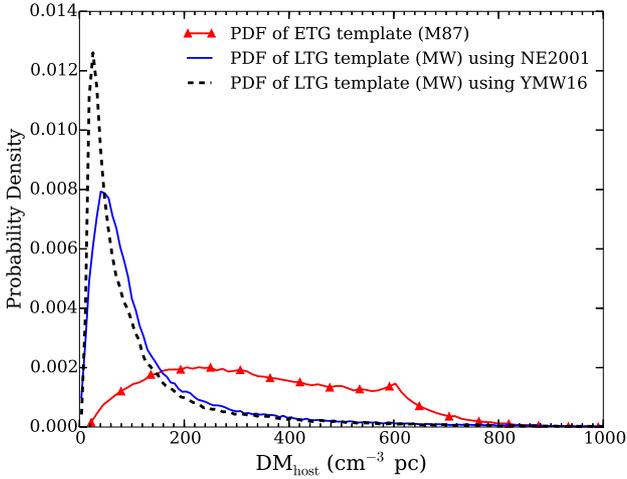}
	\caption{The DM distribution of the reference galaxies, i.e. for the Milky Way 
	and the M87. The solid curve with triangle marks (red) is PDF of DMs of
	FRBs in the M87. The solid curve (blue) and the dashed curve (black) are 
	for the Milky Way using the NE2001 model and the YMW16 model, respectively.}
	\label{fig:dmtemp} \end{figure}

We calculate the DM PDF of FRBs in the M87 using the Monte Carlo method.  First, we 
create the a million artificial FRBs with positions according to the Young 
profile \citep{Young76AJ}
\begin{equation}
	\rho_s=\rho_0\frac{\exp\left[-b(R/R_{\rm e})^{1/4}\right]}{(R/R_{\rm
	e})^{7/8}}\,, \label{eq:young} \end{equation}
where $\rho_0$ is the stellar density and $R_{\rm e}=7.7\,{\rm kpc}$ for M87
\citep{ZMS93MN}. We then compute the DM
PDF of those FRBs by integrating the electron density (i.e.  \EQ{eq:ne_etg})
along the path in random directions uniformly distributed over the full-sky $4\pi$
solid angle. The DM PDF is plotted in \FIG{fig:dmtemp}.

The DM distribution function of M87 is rather flat, due to the
spherical electron density distribution. The FRBs in ETGs is
concentrated around the galaxy centre, because of the rather compact
Young profile. For the case of M87, the little spike in the DM
distribution function peaking around 600 $\cmpc$ is due to such a
concentration. M87 is a giant elliptical galaxy, the high
DM value with a few hundred $\cmpc$ comes as no surprise. For most of the
ETGs, we expect that the DM will be much lower, because of their smaller
sizes and lower $\Ha$ luminosities.

\begin{figure*} \centering \includegraphics[width=7.in]{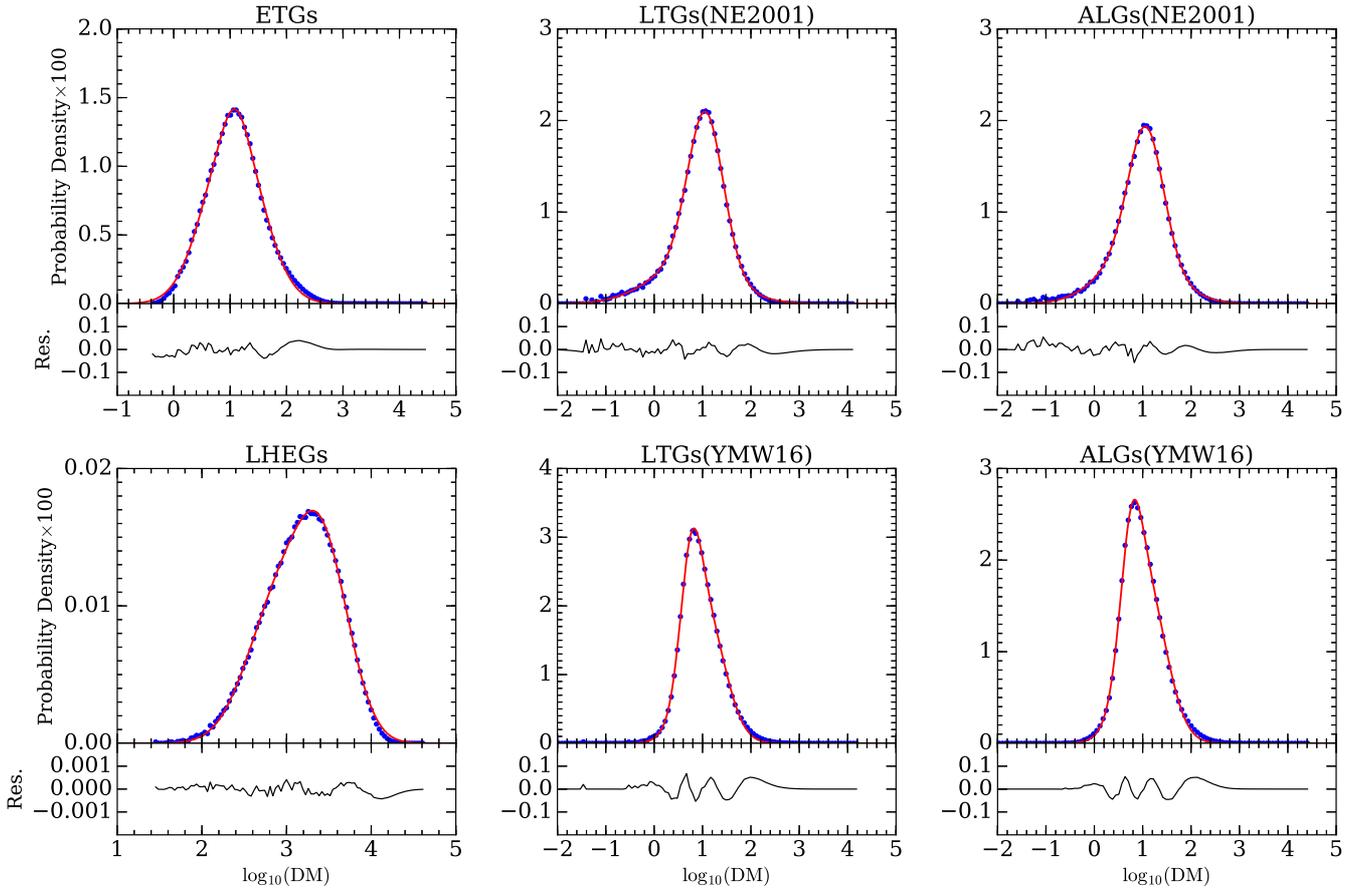}
	\caption{The DM distribution functions and the fitted functions before normalisation. Here are six galaxy categories: early-type 
	galaxies (ETGs), luminous $\Ha$ elliptical galaxies (LHEGs), late-type 
	galaxies (LTGs), and all galaxies (ALGs) using two different electron density 
	models: NE2001 and YMW16. The recipe to compute the curves is in 
	\SEC{sec:dms}. For each panel, the galaxy type and electron density model are 
	given in the title. The simulated DM distribution function using
	Monte Carlo method is plotted in blue dots. The fitted curve is plot as the 
	red curves, with residuals in the bottom panels.
} \label{fig:fits} \end{figure*}

Fixing the M87 as the reference galaxy, we compute the DM distribution function 
for all ETGs with another Monte Carlo simulation. In each step, we draw one 
sample of DM value from the M87 distribution, $\Ha$ luminosity from luminosity 
function, and $R_{\rm e}$ via $r$-band luminosity function. Then, we use 
\EQ{eq:scaldm} to compute the DM of the simulated galaxy. We repeat the 
procedures and produce the DM distributions of ETGs and LHEGs, which are plotted 
in \FIG{fig:fits}.

For reference purposes, we approximate the DM distribution using an analytical 
form. We note that the double-Gaussian function, 
i.e. 
\begin{equation} f_{\cal D}({\rm DM}_{\rm host,0})\, \D {\rm DM}_{\rm host,0}  = 
	\sum_{i=1}^{2} a_{i}
	e^{-\left(\frac{\log_{10} {\rm DM}_{\rm host,0}-b_i}{c_i}\right)^2}\, \D \DM_{\rm host,0} 
	\,.	\label{eq:dmfit}
\end{equation} 
can fit the curves rather well. The fitted parameters and the curves for those ensemble DM 
distribution functions are listed in \TAB{tab:fits} and shown in \FIG{fig:fits} 
respectively.

\begin{table*}
	\caption{The fitted parameters of DM PDF}
	\begin{center} \begin{tabular}{ccccccc}
		\hline \hline
		Parameters & ETGs & LHEGs & LTGs(NE2001) & LTGs(YMW16) & ALGs(NE2001) & ALGs(YMW16) 
		
		\\ \hline
		$a_1 (\times10^{-3})$ & 1.963 & 0.1182 & 14.31 & 17.51 & 4.899 & 13.79  \\
		$b_1$ & 1.099 & 3.441 & 1.062 & 0.759 & 0.8665 & 0.7597  \\
		$c_1$ & 0.2965 & 0.4407 & 0.5202 & 0.3013 & 1.009 & 0.3082 \\ 
		$a_2 (\times10^{-3})$ & 14.28 & 0.09462 & 3.471 & 21.19 & 12.56 & 19.96 \\ 
		$b_2$ & 1.055 & 2.906 & 0.7227 & 1.042 & 1.069 & 1.048 \\ 
		$c_2$ & 0.7262 & 0.5317 & 1.151 & 0.5791 & 0.5069 & 0.6025  \\ 
		\hline
	\end{tabular}
	\end{center} \label{tab:fits} \end{table*}

{\bf \emph{LTGs:}}  We adopt Milky Way as the reference galaxy for the LTGs.  
The Milky-Way stellar structure can be well
approximated by the combination of a thin exponential disk and a Young-profile
spheroid. We use the BS model
\citep{BS80ApJS, BS84ApJS, Bahcall86AR}
and the Young profile, i.e. \EQ{eq:young}, to model the
stellar distribution. The stellar distribution of the disk component is
\begin{equation} \rho_{\rm D}(\varpi, Z)=\rho_{\rm D0}\exp{\left[-\frac{Z}{H^*}
	\right]}\exp\left[ -\frac{\varpi-\varpi_0}{\varpi^*}\right]\,, \label{dendisk}
\end{equation} where the radial distance to the $Z$-axis is
$\varpi=\sqrt{X^2+Y^2}$. The central stellar density $\rho_{\rm
D0}=0.13\ \mathrm{pc}^{-3}$, $\varpi_0=8\ \mathrm{kpc}$, scale height $H^*=0.3\
\mathrm{kpc}$, and scale radius $\varpi^*=3.5\ \mathrm{kpc}$.  For the spheroid
component, the density profile is described by the Young profile as in
\EQ{eq:young}, with $\rho_0=2.6\times10^{-4}\,\mathrm{pc}^{-3}$, $R_{\rm e}=2.7\, 
\mathrm{kpc}$, and
$b\simeq7.7$.

The electron
density models we used are the NE2001 and the YMW16.  The simulated
DM distribution for Milky Way is plotted in
\FIG{fig:dmtemp}.  The DM distribution function of the LTG template is relatively compact 
compared to the case of the ETG template, because LTGs have an exponential drop of the 
stellar distribution and the electron density distribution in both radial and
vertical directions of the disk component. The most probable DM values are
40 and 25 $\cmpc$ for the NE2001 and the YMW16 model, respectively, which are a 
factor of 3 to 4
smaller than previous estimations
\citep{Thornton13Sci}. Using the same method described for ETGs, we compute the 
ensemble
distribution functions of LTGs. The DM distribution is shown in \FIG{fig:fits}, 
and the fitted results are in \TAB{tab:fits}.

{\bf \emph{All galaxies:}} We can combine the ETGs and LTGs to form the total 
galaxy population and define the sample as ``all galaxies'' (ALGs). 
The integrals of the \textit{r}-band luminosity functions 
\citep{Nakamura03AJ} produce the fraction number of ETGs and LTGs, which
are 23.7\% and 76.3\%, respectively.  
Due to the dominance of LTGs, the DM distribution function of ALGs is
very similar to that of LTGs.  The results are shown in \FIG{fig:fits} and 
\TAB{tab:fits}. 

With the $\DMh$ distribution, the most probable isotropic luminosity and energy 
of FRB emission can be estimated as the byproducts. The technique is described in
\APP{app:probl} and the results are given in \TAB{tab:frbs}.

\subsection{Posterior sampling and algorithm verification}
\label{sec:post}
Our likelihood is given in \EQ{eq:likfun}. Choosing the uniform prior for the 
dimensionless parameters and the uniform prior in logarithmic scale for the 
parameters with units introduces the least amount of prior information 
\citep{Gregory2005}.  We thus choose uniform prior for $\alpha$ and $\log L^*$.  
However, as we will show, we can not measure the lower cutoff of FRB luminosity 
$L_0$ yet, due to the limited FRB sample. The standard trick to determine the
upper limit \citep{LTM16} is to use the uniform prior for $L_0$. 

Instead of a direct evaluation for the integration in \EQ{eq:bayest},
the posterior calculation is usually performed using sampling
techniques. In this paper, we use the \textsc{multinest} algorithm 
\citep{Feroz09MN}, which is widely applied in astronomical applications.  
The nested sampling \citep{Skilling04}
is a Monte Carlo method to compute Bayesian evidence efficiently
and produce the posterior samples. This is done by converting the
parameter space to a set of nested shells with equal posterior
values and iteratively sampling with replacements in the nested
volume. To achieve a better efficiency, \textsc{multinest} further partitions
the nested samples. In our posterior sampling, we use the python language 
interface \textsc{pymultinest} 
\footnote{https://johannesbuchner.github.io/PyMultiNest/} when calling the
\textsc{multinest} library.

We test the likelihood function, prior choice, and \textsc{multinest} sampler 
using the simulated mock data set. The mock data is generated by (1) sampling the 
luminosity of FRB according to the input FRB luminosity function; (2) sampling 
the FRB redshift according to \EQ{eq:fz}; (3) sampling the host galaxy DM 
for fixed galaxy type according to the distribution function in \EQ{eq:dmfit}; (4) sampling the local DM according to uniform probability distribution mentioned in \SEC{sec:dislocdm};
(5) sampling beam response according to the distribution mentioned in 
\SEC{sec:beam}; (6) calculating the FRB flux density and extragalactic DM; and 
(7) selecting the sources above the detection threshold. 

\begin{figure}
\centering
\subfloat[Posterior distribution using uniform prior for $\log L_0$]
{\includegraphics[width=3.5in]{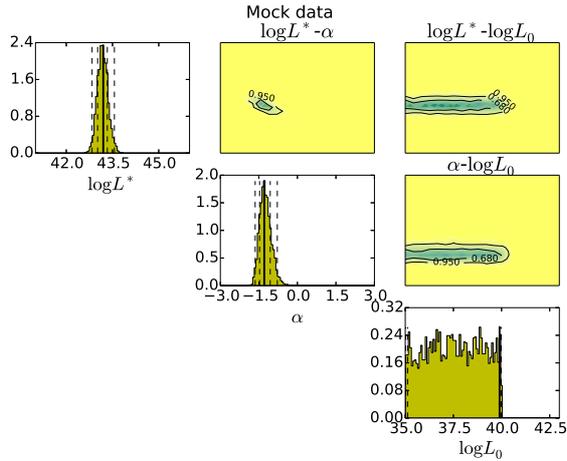}}\\
\subfloat[Posterior distribution using uniform prior for $L_0$]
{\includegraphics[width=3.5in]{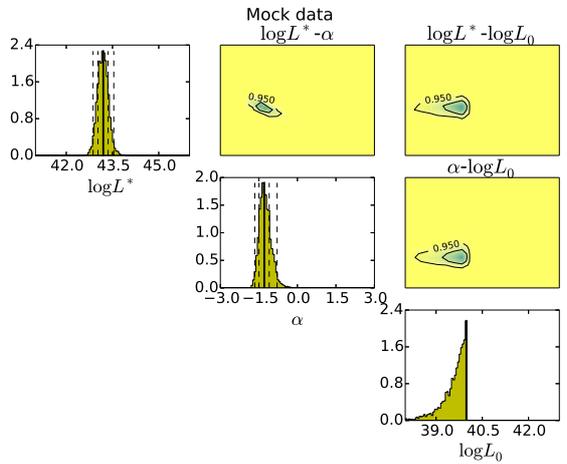}}
\caption{Inference for the mock data. The parameters used in creating the mock 
data are $\log L^*=43.0$, $\alpha=-1.0$, and $\log L_0=39.0$. The diagonal histogram 
is the marginalised one-dimensional posterior distribution for each of the parameters. For $\log L^*$ and $\alpha$,
 the solid lines denote the most probable parameter value, while
the dashed lines indicate the 67\% and the 95\% confidence level. 
For $\log L_0$, the solid line denote the upper limit value with 95\% confidence level.
The off-diagonal contour plots are for the marginalised two-dimensional posteriors, 
with parameters indicated in the title. The inner and outer black contours are for 67\% and 95\% 
confidence levels. In the panel (\textbf{a}), we adopt uniform prior for $\log L_0$.
As indicated by the flat histogram of the $\log L_0$ distribution, we can not get 
good measurement for the value of $L_0$. This is mainly due to the limited 
number sample (100 FRBs are simulated here).  We switch the uniform prior for 
$L_0$ to produces the upper limit of $L_0$, as shown in the panel ({\bf b}).}
\label{fig:mocdat}
\end{figure}

The results from analysing the mock data are shown in \FIG{fig:mocdat}. As one 
can see, the current Bayesian inference recovers the parameters of the input 
luminosity function rather well.

\section{Results for the FRB luminosity function}
\label{sec:res}
We perform our Bayesian inference and use the data of 33 FRBs to measure the FRB 
luminosity functions for the following six cases.

{\bf Case 1, ETG-NE2001} Host galaxy is assumed as ETGs, and NE2001 electron 
model is assumed for Galactic DM correction.

{\bf Case 2, ETG-YMW16} The same as the {\bf Case 1}, except using YMW16 to 
correct the Galactic DM.

{\bf Case 3, LTG-NE2001} The same as the {\bf Case 1}, except that the host 
galaxy is assumed as LTGs.

{\bf Case 4, LTG-YMW16} The same as the {\bf Case 2}, except that the host 
galaxy is assumed as LTGs.

{\bf Case 5, ALG-NE2001} The same as the {\bf Case 1}, except that the host 
galaxy is assumed as the composition of both LTGs and ETGs.

{\bf Case 6, ALG-YMW16} The same as the {\bf Case 2}, except that the host 
galaxy is assumed as the composition of both LTGs and ETGs.

\begin{figure*}
\centering
\includegraphics[width=7in]{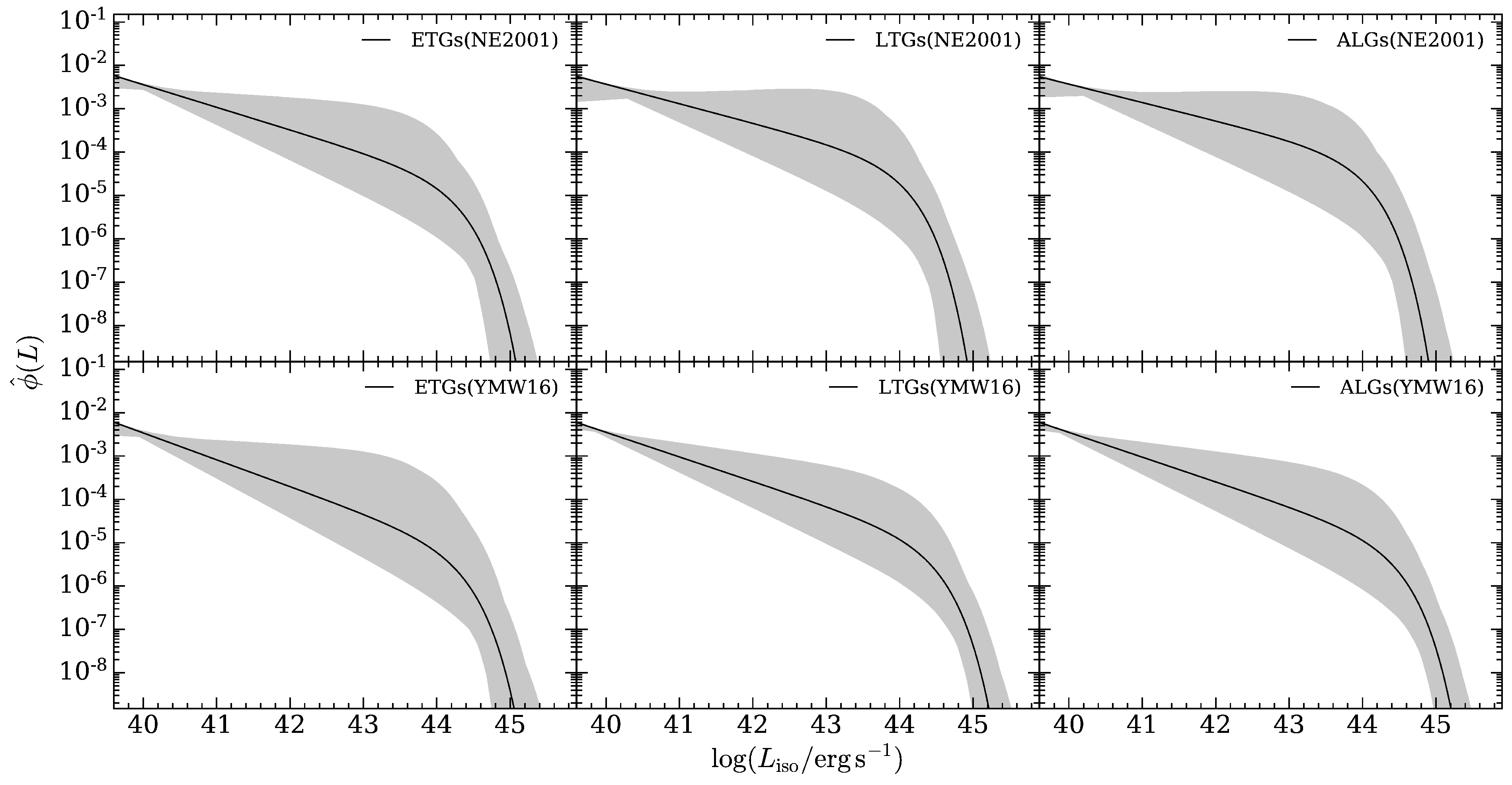}
\caption{The FRB normalised luminosity functions.  In each panel, the solid line 
(black) is the best fitting luminosity function, and the shaded area (grey) 
shows the luminosity function within 1$\sigma$ error. The meaning of the labels 
are, $1^{\circ}$ ETGs(NE2001): Galactic foreground removed with NE2001 and 
assuming ETG as the
host galaxy; $2^{\circ}$ ETGs(YMW16): the same as $1^{\circ}$ but with galactic 
foreground removed with YMW16; $3^{\circ}$ LTGs(NE2001): Galactic foreground 
removed with NE2001 and assuming LTG
as the host galaxy; $4^{\circ}$ LTGs(YMW16): the same as $3^{\circ}$ but with 
galactic foreground removed with YMW16; $5^{\circ}$ ALGs(NE2001): Galactic 
foreground removed with NE2001 and assuming mixed galaxy (ALG)
as the host galaxy; $6^{\circ}$ ALGs(YMW16): the same as $5^{\circ}$ but with 
galactic foreground removed with YMW16.}
\label{fig:frblf}
\end{figure*}

The plots for posterior PDF of all six cases are summarised in \APP{sec:postes}.  
The maximal likelihood estimators and the errors are summarised in 
\TAB{tab:frblf}. For each of the six cases, we compared two scenarios, i.e.  
removing the Galactic halo contributions $\DMd$ or not in the pre-processing 
stage. The shapes of luminosity functions together with the confidence regions 
are plotted in \FIG{fig:frblf}.  Interestingly, even though the DM distribution 
functions depend on the galaxy types, the inferred luminosity functions do not 
vary much, where
the power-law index $\alpha\simeq -1.5$ and cut-off luminosity $\log L^*\simeq 
44.2$. We can not measure the low cut-off luminosity $\log L_0$ due to the limited 
number of currently known FRBs, however the 95\%-confidence-level upper limit 
$\log L_0\le 41.9$ is derived with a uniform prior for $L_0$.

\begin{table*}
\caption{The parameters of FRB luminosity function}
\begin{center}
\begin{tabular}{ccccccc}
\hline
\hline
Galaxy type & \multicolumn{3}{c}{No modelling for Galactic halo} & 
\multicolumn{3}{c}{Removed Galactic halo}\\ \cmidrule(lr){2-4} 
\cmidrule(lr){5-7}
\cmidrule(lr){2-4} \cmidrule(lr){5-7}
& $\alpha \, (1\sigma)$ &  $\log L^* \, (1\sigma)$ & $\log L_0$ (95\% C.L.) & $\alpha \,
(1\sigma)$ &  $\log L^* \, (1\sigma)$ & $\log L_0$ (95\% C.L.) \\ 
\hline
ETGs (NE2001) & $-1.52^{+0.24}_{-0.23}$ & $44.14^{+0.23}_{-0.33}$ & $\le41.75$
 & $-1.57^{+0.19}_{-0.26}$ & $44.10^{+0.23}_{-0.33}$ & $\le41.56$\\
 
ETGs (YMW16) & $-1.62^{+0.29}_{-0.21}$ & $44.18^{+0.26}_{-0.38}$ & $\le41.96$ 
& $-1.67^{+0.21}_{-0.25}$ & $44.23^{+0.27}_{-0.38}$ & $\le41.82$\\

LTGs (NE2001) & $-1.45^{+0.31}_{-0.28}$ & $43.94^{+0.22}_{-0.35}$ & $\le41.74$
 & $-1.50^{+0.25}_{-0.26}$ & $43.87^{+0.27}_{-0.30}$ & $\le41.56$\\
 
LTGs (YMW16) & $-1.57^{+0.17}_{-0.22}$ & $44.32^{+0.22}_{-0.24}$ & $\le41.96$
 & $-1.60^{+0.15}_{-0.19}$ & $44.29^{+0.33}_{-0.20}$ & $\le41.82$ \\
 
ALGs (NE2001) & $-1.42^{+0.27}_{-0.27}$ & $43.90^{+0.30}_{-0.29}$ & $\le41.74$
 & $-1.51^{+0.26}_{-0.25}$ & $43.89^{+0.26}_{-0.28}$ & $\le41.56$ \\
 
ALGs (YMW16) & $-1.57^{+0.19}_{-0.21}$ & $44.31^{+0.22}_{-0.27}$ & $\le41.96$
 & $-1.63^{+0.16}_{-0.19}$ & $44.34^{+0.21}_{-0.29}$ & $\le41.82$ \\ 
\hline
\end{tabular}
\end{center}

\label{tab:frblf}
\end{table*}

\section{Discussion}
\label{sec:disc}

In this paper, we measured the FRB luminosity function using the Bayesian method 
under different assumptions for the host galaxy type. The Bayesian method helped 
dealing with the missing information, i.e. the distances of FRBs and beam 
response, which are difficult
 to handle otherwise. Assuming the Schechter form for the luminosity 
 function, we measured the power law index and high cut-off luminosity, where 
 $\alpha\simeq-1.5$, and $L^*\simeq10^{44}\,\ergs$. As byproducts, we also used 
 the Bayesian method (see \APP{app:probl}) to infer the most-probable redshift, 
 isotropic luminosity and energy of each source with the values listed in 
 \APP{app:dattab}. 

The FRB luminosity power-law indices, we measured, range from $-1.8$ to $-1.2$. Such 
values also agree with (i) the power-law indices of pulsars' giant pulse flux 
distribution at lower frequency (--1.7, \citealt{KSL12}); (ii) the mean power-law 
indices of radio emission of pulsars ($-1.65$ to $-2.2$, \citealt{HWX16, JVK18}); (iii) the power-law 
index of luminosity function of long gamma-ray bursts ($-1.3$ to $-2.3$;
\citealt{SZL15,PGS16}) (iv) short gamma-ray bursts ($-1.5$ to $-1.7$, \citealt{SZL15}); (v) compact binary mergers ($-1.2$ to $-1.7$, \citealt{CYZ18}).
We can not pin down the radiation mechanisms based on the 
FRB luminosity function. However, the similarity between it and those of
of other astrophysical sources may suggest a common underlying rule of 
defining burst populations of different kinds.

The distance information of FRBs is determined from the DM values.  We modelled 
the DM from three major contributions, i.e. from the Milky way, the IGM, and the 
FRB host galaxy. We also compared the results to evaluate the effects of Galaxy 
halo contribution.  We showed that the parameters for the luminosity function 
are rather insensitive to the modeling details.

We modelled the electron density distribution functions for two major cases 
in the paper, i.e. spiral galaxies and elliptical galaxies. The most 
likely values of $\DMh$ for these two cases are approximately 10 and 15 $\cmpc$, respectively.  
Such host galaxy DM values are at least one order of magnitude smaller than that 
of the IGM contribution. Although the most uncertain part in our modelling is 
the $\DMh$ distribution, the parameters inference for the luminosity function is 
rather robust as $\DMe\gg\DMh$.  We can tolerate  the missing information 
such as the the $\Ha$ filling
factor, the stellar distribution in galaxies, halo DM or FRB source DM. The
characteristic host galaxy DM values we estimated are less than 100 $\cmpc$.
Despite this, considering the scattering of the distribution, our results are 
still compatible with the values estimated before \citep{Thornton13Sci, XH15RAA, 
Yang17ApJ}.  

The average DM value of ETGs we calculated here is higher than that for
LTGs. This is mainly due to the stellar distribution and galaxy
morphology.  The concentration of FRBs in the central region of ETGs
produce higher value of DM for the ETGs than for the LTG, where a lower scale 
height of LTGs leads to a lower DM.

For the case of LHEGs, i.e. elliptical galaxies with $\Ha$ luminosity 
function in \citet{Nakamura04AJ}, the most likely DM host is
$\DMh\simeq3000\,\cmpc$. 
Considering that the observed $\DMh$ is smaller by a factor 
$(1+z)$ and the roughly linear increase of $\DMi$ with 
redshift, one obtains that an FRB with $z>2$ would have a $\DMe$
exceeding $2750\,\cmpc$ \citep{Zhang18arXiv} which is larger 
than the maximum observed $\rm DM_E$ (e.g. $2583.1\,\cmpc$ from 
FRB~160102, \citealt{Bhandari18MN}). 
If FRBs all have LHEG hosts, the probability of detecting one FRB 
with $\DMe\le2750$ is only $\simeq35\%$, as computed by integrating $\fd(\DMh)$ from 
$0$ to $2750$. Thus there is only a miniscule chance  ($8\times10^{-16}$) to observe all 33 
FRBs with $\DMe\le 2750$~cm$^{-3}$~pc. We conclude that it is unlikely that the LHEGs are the 
host galaxies for FRBs, unless all FRBs lie around the galaxy outskirts if they originate in LHEGs.

The $\DMh$ distribution function of all the galaxies enables us to
infer the corresponding isotropic luminosity and energy of FRB
emission as listed in \TAB{tab:frbs}. Using only DM as the distance indicator, our 
inferred most probable redshift of FRB 121102 ranges from 0.198 to 0.271 at a
$2\sigma$ confidence level. This
is roughly consistent with the true redshift 0.193 \citep{Tendulkar17ApJ}.
The slightly higher value of the inferred redshift may be resulted from the long tail of PDF for $\DMh$ as computed in \SEC{sec:dms}.
The excess may also come from an underestimate of the electron density in the Milky Way
halo or in the FRB environment. Alternatively, it could be due to the deviation
of the mean cosmological $\DMi$ due to the existence of large scale structures,
so that the line of sight towards FRB 121102 may have pieced through an over-dense
region in the IGM. We expect that more FRBs with optically-measured
redshift will help us to clarify these issues.

We used M87 and Milky Way as the template galaxies in this
study. The choice is made because they are the only two galaxies of each type
that previously have accurate measurements on both electron density 
distributions and
$\Ha$ luminosities. As a caveat, both galaxies may not be the typical
example of ETGs or LTGs. The M87 is a giant elliptical galaxy, and
the Milky way has a relatively low gas fraction \citep{KE2012AR}. We
can still use Milky Way and M87 as the reference values, thanks to our 
scaling method, which accounts for the galaxy size, electron density, and star 
formation history evolution.

We assumed that the FRB distribution in the galaxies follows the
stellar distribution. In Milky Way, the steller distribution has
low scale height than that of the neutron star distribution. If the
FRBs are of a magnetar or pulsar origin, the host galaxy DM may be slightly
overestimated here. However, since the host galaxy DM is already smaller than 
the observed DM, such a second-order small perturbation
can be well neglected without affecting the luminosity function
inference.

We modelled the luminosity distribution using the Schechter function.
The measured cut-off luminosity $\log L^*\simeq44.2\,\ergs$ with
an error of 0.3 dex indicates that the simple power-law
distribution is not precise enough at the high luminosity end. This 
also helps planning future FRB surveys.  For FRBs with distances of $\sim$ 1 Gpc, 
the size of a radio telescope for FRB survey
should be at least 10 meters to get ${\rm S/N}\ge 7$.

The possible off-centre position of an FRB in the main beam, without 
modelling, leads to an underestimate for the FRB luminosity 
\citep{Niino18ApJ}.  We include such an uncertainty through the Bayesian 
marginalisation. It turns out that the difference in the parameters of inferred 
luminosity function is not significant between the two cases, regardless of whether 
or not the beam response is taken into account.
Without modelling the beam response, the power-law  index of Schechter function 
is slightly flatter, and the cut-off luminosity is relatively lower. Taking the 
case of ALGs-YMW16 as an example, we get  $\alpha=-1.56^{+0.21}_{-0.20}$ and 
$\log L^*=44.19^{+0.22}_{-0.24}$ with no beam response modelling, whereas
$\alpha=-1.57^{+0.19}_{-0.21}$ and $\log L^*=44.31^{+0.22}_{-0.27}$ with beam 
response marginalisation. As the difference is still within 1$\sigma$ 
confidence level, we conclude that the beam response plays a limited role in parameter inference 
for the current limited sample of FRBs.

We could only obtain the upper limits of the lower cut-off luminosity,
i.e. $L_0$, due to the limited sample of FRBs (\TAB{tab:frblf}).
The current upper limit of $\log L_0<42.0$ is not capable of
testing the FRB model yet. In order to measure the true value of $L_0$,
a large number of nearby FRBs are required. 

Due to the
unknown spectral shape and width, our luminosity function is based on the
reference bandwidth of 1 GHz. This is motivated by the observations of the 
repeating FRB 121102,
which indicates a $\sim$ 1 GHz bandwidth \citep{GSP18}. In general, the 
parameter $L^*$ in the luminosity function scales with the reference bandwidth.  
The assumption of a $\sim$ 1 GHz bandwidth can be revised later.  Little information is 
available for the spectrum of FRBs at present, and scintillation may introduce 
a strong bias in determining the true bandwidth.  The measurement in this paper 
can be further improved, if future observations will provide more information.
We expect that the upcoming large field-of-view facilities,
e.g. CHIME\citep{Ng17CHIME}, ASKAP\citep{Macquart10PASA},
MeerKAT\citep{Booth12AfrSk} and instruments with higher sensitivity,
e.g.  ALFABURST survey \citep{Foster17ALFABURST}, FAST \citep{Nan11},
SKA \citep{Macquart15SKA},  and QTT \citep{Wang17}, will provide
more opportunities to detect more nearby FRBs and reveal the details of the FRB 
spectra.

\section*{Acknowledgements}
This work was supported by NSFC U15311243 National Basic Research Program
of China, 973 Program, 2015CB857101, XDB23010200, 11690024, 11373011, and 
funding from TianShanChuangXinTuanDui and the Max-Planck Partner Group.  We are 
grateful to Luis C. Ho, Joris P. W. Verbiest and Yuan-pei Yang for reading 
through the paper and giving their helpful suggestions and comments.

\newpage



\bibliographystyle{mnras}
\bibliography{ms} 

\begin{thebibliography}{}
\makeatletter
\relax
\def\mn@urlcharsother{\let\do\@makeother \do\$\do\&\do\#\do\^\do\_\do\%\do\~}
\def\mn@doi{\begingroup\mn@urlcharsother \@ifnextchar [ {\mn@doi@}
  {\mn@doi@[]}}
\def\mn@doi@[#1]#2{\def\@tempa{#1}\ifx\@tempa\@empty \href
  {http://dx.doi.org/#2} {doi:#2}\else \href {http://dx.doi.org/#2} {#1}\fi
  \endgroup}
\def\mn@eprint#1#2{\mn@eprint@#1:#2::\@nil}
\def\mn@eprint@arXiv#1{\href {http://arxiv.org/abs/#1} {{\tt arXiv:#1}}}
\def\mn@eprint@dblp#1{\href {http://dblp.uni-trier.de/rec/bibtex/#1.xml}
  {dblp:#1}}
\def\mn@eprint@#1:#2:#3:#4\@nil{\def\@tempa {#1}\def\@tempb {#2}\def\@tempc
  {#3}\ifx \@tempc \@empty \let \@tempc \@tempb \let \@tempb \@tempa \fi \ifx
  \@tempb \@empty \def\@tempb {arXiv}\fi \@ifundefined
  {mn@eprint@\@tempb}{\@tempb:\@tempc}{\expandafter \expandafter \csname
  mn@eprint@\@tempb\endcsname \expandafter{\@tempc}}}

\bibitem[\protect\citeauthoryear{{Bahcall}}{{Bahcall}}{1986}]{Bahcall86AR}
{Bahcall} J.~N.,  1986, \mn@doi [\araa] {10.1146/annurev.aa.24.090186.003045},
  \href {http://adsabs.harvard.edu/abs/1986ARA%26A..24..577B} {24, 577}

\bibitem[\protect\citeauthoryear{{Bahcall} \& {Soneira}}{{Bahcall} \&
  {Soneira}}{1980}]{BS80ApJS}
{Bahcall} J.~N.,  {Soneira} R.~M.,  1980, \mn@doi [\apjs] {10.1086/190685},
  \href {http://adsabs.harvard.edu/abs/1980ApJS...44...73B} {44, 73}

\bibitem[\protect\citeauthoryear{{Bahcall} \& {Soneira}}{{Bahcall} \&
  {Soneira}}{1984}]{BS84ApJS}
{Bahcall} J.~N.,  {Soneira} R.~M.,  1984, \mn@doi [\apjs] {10.1086/190948},
  \href {http://adsabs.harvard.edu/abs/1984ApJS...55...67B} {55, 67}

\bibitem[\protect\citeauthoryear{{Bannister} et~al.,}{{Bannister}
  et~al.}{2017}]{Bannister17ApJ}
{Bannister} K.~W.,  et~al., 2017, \mn@doi [\apjl] {10.3847/2041-8213/aa71ff},
  \href {http://adsabs.harvard.edu/abs/2017ApJ...841L..12B} {841, L12}

\bibitem[\protect\citeauthoryear{{Bassa} et~al.,}{{Bassa}
  et~al.}{2017}]{Bassa17ApJ}
{Bassa} C.~G.,  et~al., 2017, \mn@doi [\apjl] {10.3847/2041-8213/aa7a0c}, \href
  {http://adsabs.harvard.edu/abs/2017ApJ...843L...8B} {843, L8}

\bibitem[\protect\citeauthoryear{{Bhandari} et~al.,}{{Bhandari}
  et~al.}{2018}]{Bhandari18MN}
{Bhandari} S.,  et~al., 2018, \mn@doi [\mnras] {10.1093/mnras/stx3074}, \href
  {http://adsabs.harvard.edu/abs/2018MNRAS.475.1427B} {475, 1427}

\bibitem[\protect\citeauthoryear{{Binggeli}, {Sandage}  \&
  {Tammann}}{{Binggeli} et~al.}{1988}]{BST88}
{Binggeli} B.,  {Sandage} A.,   {Tammann} G.~A.,  1988, \mn@doi [\araa]
  {10.1146/annurev.aa.26.090188.002453}, \href
  {http://adsabs.harvard.edu/abs/1988ARA%26A..26..509B} {26, 509}

\bibitem[\protect\citeauthoryear{{Bonetti}, {Ellis}, {Mavromatos}, {Sakharov},
  {Sarkisyan-Grinbaum}  \& {Spallicci}}{{Bonetti} et~al.}{2016}]{Bonetti16PLB}
{Bonetti} L.,  {Ellis} J.,  {Mavromatos} N.~E.,  {Sakharov} A.~S.,
  {Sarkisyan-Grinbaum} E.~K.,   {Spallicci} A.~D.~A.~M.,  2016, \mn@doi
  [Physics Letters B] {10.1016/j.physletb.2016.04.035}, \href
  {http://adsabs.harvard.edu/abs/2016PhLB..757..548B} {757, 548}

\bibitem[\protect\citeauthoryear{{Bonetti}, {Ellis}, {Mavromatos}, {Sakharov},
  {Sarkisyan-Grinbaum}  \& {Spallicci}}{{Bonetti} et~al.}{2017}]{Bonetti17PLB}
{Bonetti} L.,  {Ellis} J.,  {Mavromatos} N.~E.,  {Sakharov} A.~S.,
  {Sarkisyan-Grinbaum} E.~K.,   {Spallicci} A.~D.~A.~M.,  2017, \mn@doi
  [Physics Letters B] {10.1016/j.physletb.2017.03.014}, \href
  {http://adsabs.harvard.edu/abs/2017PhLB..768..326B} {768, 326}

\bibitem[\protect\citeauthoryear{{Booth} \& {Jonas}}{{Booth} \&
  {Jonas}}{2012}]{Booth12AfrSk}
{Booth} R.~S.,  {Jonas} J.~L.,  2012, African Skies, \href
  {http://adsabs.harvard.edu/abs/2012AfrSk..16..101B} {16, 101}

\bibitem[\protect\citeauthoryear{{Born} \& {Wolf}}{{Born} \&
  {Wolf}}{1999}]{BWOptics}
{Born} M.,  {Wolf} E.,  eds, 1999, {Principles of optics : electromagnetic
  theory of propagation, interference and diffraction of light}.
Cambridge University Press, New York

\bibitem[\protect\citeauthoryear{{Bregman} \& {Lloyd-Davies}}{{Bregman} \&
  {Lloyd-Davies}}{2007}]{BL07}
{Bregman} J.~N.,  {Lloyd-Davies} E.~J.,  2007, \mn@doi [\apj] {10.1086/521321},
  \href {http://adsabs.harvard.edu/abs/2007ApJ...669..990B} {669, 990}

\bibitem[\protect\citeauthoryear{{Burke-Spolaor} \&
  {Bannister}}{{Burke-Spolaor} \& {Bannister}}{2014}]{Burke-Spolaor14ApJ}
{Burke-Spolaor} S.,  {Bannister} K.~W.,  2014, \mn@doi [\apj]
  {10.1088/0004-637X/792/1/19}, \href
  {http://adsabs.harvard.edu/abs/2014ApJ...792...19B} {792, 19}

\bibitem[\protect\citeauthoryear{{Cai}, {Sabancilar}  \& {Vachaspati}}{{Cai}
  et~al.}{2012a}]{CSV12}
{Cai} Y.-F.,  {Sabancilar} E.,   {Vachaspati} T.,  2012a, \mn@doi [\prd]
  {10.1103/PhysRevD.85.023530}, \href
  {http://adsabs.harvard.edu/abs/2012PhRvD..85b3530C} {85, 023530}

\bibitem[\protect\citeauthoryear{{Cai}, {Sabancilar}, {Steer}  \&
  {Vachaspati}}{{Cai} et~al.}{2012b}]{CSS12}
{Cai} Y.-F.,  {Sabancilar} E.,  {Steer} D.~A.,   {Vachaspati} T.,  2012b,
  \mn@doi [\prd] {10.1103/PhysRevD.86.043521}, \href
  {http://adsabs.harvard.edu/abs/2012PhRvD..86d3521C} {86, 043521}

\bibitem[\protect\citeauthoryear{{Caleb} et~al.,}{{Caleb}
  et~al.}{2017}]{Caleb17MN}
{Caleb} M.,  et~al., 2017, \mn@doi [\mnras] {10.1093/mnras/stx638}, \href
  {http://adsabs.harvard.edu/abs/2017MNRAS.468.3746C} {468, 3746}

\bibitem[\protect\citeauthoryear{{Cao}, {Yu}  \& {Dai}}{{Cao}
  et~al.}{2017}]{Cao17ApJ}
{Cao} X.-F.,  {Yu} Y.-W.,   {Dai} Z.-G.,  2017, \mn@doi [\apjl]
  {10.3847/2041-8213/aa6af2}, \href
  {http://adsabs.harvard.edu/abs/2017ApJ...839L..20C} {839, L20}

\bibitem[\protect\citeauthoryear{{Cao}, {Yu}  \& {Zhou}}{{Cao}
  et~al.}{2018}]{CYZ18}
{Cao} X.-F.,  {Yu} Y.-W.,   {Zhou} X.,  2018, \mn@doi [\apj]
  {10.3847/1538-4357/aabadd}, \href
  {http://adsabs.harvard.edu/abs/2018ApJ...858...89C} {858, 89}

\bibitem[\protect\citeauthoryear{{Cavagnolo}, {Donahue}, {Voit}  \&
  {Sun}}{{Cavagnolo} et~al.}{2009}]{Cavagnolo09ApJS}
{Cavagnolo} K.~W.,  {Donahue} M.,  {Voit} G.~M.,   {Sun} M.,  2009, \mn@doi
  [\apjs] {10.1088/0067-0049/182/1/12}, \href
  {http://adsabs.harvard.edu/abs/2009ApJS..182...12C} {182, 12}

\bibitem[\protect\citeauthoryear{{Champion} et~al.,}{{Champion}
  et~al.}{2016}]{Champion16MN}
{Champion} D.~J.,  et~al., 2016, \mn@doi [\mnras] {10.1093/mnrasl/slw069},
  \href {http://adsabs.harvard.edu/abs/2016MNRAS.460L..30C} {460, L30}

\bibitem[\protect\citeauthoryear{{Chatterjee} et~al.,}{{Chatterjee}
  et~al.}{2017}]{Chatterjee17Nat}
{Chatterjee} S.,  et~al., 2017, \mn@doi [\nat] {10.1038/nature20797}, \href
  {http://adsabs.harvard.edu/abs/2017Natur.541...58C} {541, 58}

\bibitem[\protect\citeauthoryear{{Chennamangalam}, {Lorimer}, {Mandel}  \&
  {Bagchi}}{{Chennamangalam} et~al.}{2013}]{CLM13}
{Chennamangalam} J.,  {Lorimer} D.~R.,  {Mandel} I.,   {Bagchi} M.,  2013,
  \mn@doi [\mnras] {10.1093/mnras/stt205}, \href
  {http://adsabs.harvard.edu/abs/2013MNRAS.431..874C} {431, 874}

\bibitem[\protect\citeauthoryear{{Connor}, {Sievers}  \& {Pen}}{{Connor}
  et~al.}{2016}]{Connor16MN}
{Connor} L.,  {Sievers} J.,   {Pen} U.-L.,  2016, \mn@doi [\mnras]
  {10.1093/mnrasl/slv124}, \href
  {http://adsabs.harvard.edu/abs/2016MNRAS.458L..19C} {458, L19}

\bibitem[\protect\citeauthoryear{{Cordes} \& {Lazio}}{{Cordes} \&
  {Lazio}}{2002}]{CL02}
{Cordes} J.~M.,  {Lazio} T.~J.~W.,  2002, ArXiv Astrophysics e-prints, \href
  {http://adsabs.harvard.edu/abs/2002astro.ph..7156C} {}

\bibitem[\protect\citeauthoryear{{Cordes} \& {Wasserman}}{{Cordes} \&
  {Wasserman}}{2016}]{Cordes16MN}
{Cordes} J.~M.,  {Wasserman} I.,  2016, \mn@doi [\mnras]
  {10.1093/mnras/stv2948}, \href
  {http://adsabs.harvard.edu/abs/2016MNRAS.457..232C} {457, 232}

\bibitem[\protect\citeauthoryear{{Dai}, {Wang}, {Wu}  \& {Huang}}{{Dai}
  et~al.}{2016}]{Dai16ApJ}
{Dai} Z.~G.,  {Wang} J.~S.,  {Wu} X.~F.,   {Huang} Y.~F.,  2016, \mn@doi [\apj]
  {10.3847/0004-637X/829/1/27}, \href
  {http://adsabs.harvard.edu/abs/2016ApJ...829...27D} {829, 27}

\bibitem[\protect\citeauthoryear{{Deng} \& {Zhang}}{{Deng} \&
  {Zhang}}{2014}]{DZ14ApJ}
{Deng} W.,  {Zhang} B.,  2014, \mn@doi [\apjl] {10.1088/2041-8205/783/2/L35},
  \href {http://adsabs.harvard.edu/abs/2014ApJ...783L..35D} {783, L35}

\bibitem[\protect\citeauthoryear{{Dolag}, {Gaensler}, {Beck}  \&
  {Beck}}{{Dolag} et~al.}{2015}]{Dolag15MN}
{Dolag} K.,  {Gaensler} B.~M.,  {Beck} A.~M.,   {Beck} M.~C.,  2015, \mn@doi
  [\mnras] {10.1093/mnras/stv1190}, \href
  {http://adsabs.harvard.edu/abs/2015MNRAS.451.4277D} {451, 4277}

\bibitem[\protect\citeauthoryear{{Fabricant} \& {Gorenstein}}{{Fabricant} \&
  {Gorenstein}}{1983}]{FG83ApJ}
{Fabricant} D.,  {Gorenstein} P.,  1983, \mn@doi [\apj] {10.1086/160890}, \href
  {http://adsabs.harvard.edu/abs/1983ApJ...267..535F} {267, 535}

\bibitem[\protect\citeauthoryear{{Falcke} \& {Rezzolla}}{{Falcke} \&
  {Rezzolla}}{2014}]{FR14A&A}
{Falcke} H.,  {Rezzolla} L.,  2014, \mn@doi [\aap]
  {10.1051/0004-6361/201321996}, \href
  {http://adsabs.harvard.edu/abs/2014A%26A...562A.137F} {562, A137}

\bibitem[\protect\citeauthoryear{{Fan}, {Carilli}  \& {Keating}}{{Fan}
  et~al.}{2006}]{FCK06AR}
{Fan} X.,  {Carilli} C.~L.,   {Keating} B.,  2006, \mn@doi [\araa]
  {10.1146/annurev.astro.44.051905.092514}, \href
  {http://adsabs.harvard.edu/abs/2006ARA%26A..44..415F} {44, 415}

\bibitem[\protect\citeauthoryear{{Farah} et~al.,}{{Farah}
  et~al.}{2017}]{Farah17ATel}
{Farah} W.,  et~al., 2017, The Astronomer's Telegram, \href
  {http://adsabs.harvard.edu/abs/2017ATel10697....1F} {10697}

\bibitem[\protect\citeauthoryear{{Farah} et~al.,}{{Farah}
  et~al.}{2018}]{Farah18MN}
{Farah} W.,  et~al., 2018, \mn@doi [\mnras] {10.1093/mnras/sty1122}, \href
  {http://adsabs.harvard.edu/abs/2018MNRAS.478.1209F} {478, 1209}

\bibitem[\protect\citeauthoryear{{Feroz}, {Hobson}  \& {Bridges}}{{Feroz}
  et~al.}{2009}]{Feroz09MN}
{Feroz} F.,  {Hobson} M.~P.,   {Bridges} M.,  2009, \mn@doi [\mnras]
  {10.1111/j.1365-2966.2009.14548.x}, \href
  {http://adsabs.harvard.edu/abs/2009MNRAS.398.1601F} {398, 1601}

\bibitem[\protect\citeauthoryear{Fisz}{Fisz}{1963}]{Fisz63}
Fisz M.,  1963, Probability theory and mathematical statistics, 3 edn.
Hohn Wiley \& Sons, Inc, New York, NY, USA

\bibitem[\protect\citeauthoryear{{Foster} et~al.,}{{Foster}
  et~al.}{2017}]{Foster17ALFABURST}
{Foster} G.,  et~al., 2017, preprint, \href
  {http://adsabs.harvard.edu/abs/2017arXiv171010806F} {} (\mn@eprint {arXiv}
  {1710.10806})

\bibitem[\protect\citeauthoryear{{Fukugita}, {Hogan}  \& {Peebles}}{{Fukugita}
  et~al.}{1998}]{Fukugita98ApJ}
{Fukugita} M.,  {Hogan} C.~J.,   {Peebles} P.~J.~E.,  1998, \mn@doi [\apj]
  {10.1086/306025}, \href {http://adsabs.harvard.edu/abs/1998ApJ...503..518F}
  {503, 518}

\bibitem[\protect\citeauthoryear{{Gaensler}, {Madsen}, {Chatterjee}  \&
  {Mao}}{{Gaensler} et~al.}{2008}]{GM08}
{Gaensler} B.~M.,  {Madsen} G.~J.,  {Chatterjee} S.,   {Mao} S.~A.,  2008,
  \mn@doi [\pasa] {10.1071/AS08004}, \href
  {http://adsabs.harvard.edu/abs/2008PASA...25..184G} {25, 184}

\bibitem[\protect\citeauthoryear{{Gajjar} et~al.,}{{Gajjar}
  et~al.}{2018}]{GSP18}
{Gajjar} V.,  et~al., 2018, preprint, \href
  {http://adsabs.harvard.edu/abs/2018arXiv180404101G} {} (\mn@eprint {arXiv}
  {1804.04101})

\bibitem[\protect\citeauthoryear{{Gao}, {Li}  \& {Zhang}}{{Gao}
  et~al.}{2014}]{Gao14ApJ}
{Gao} H.,  {Li} Z.,   {Zhang} B.,  2014, \mn@doi [\apj]
  {10.1088/0004-637X/788/2/189}, \href
  {http://adsabs.harvard.edu/abs/2014ApJ...788..189G} {788, 189}

\bibitem[\protect\citeauthoryear{{Geng} \& {Huang}}{{Geng} \&
  {Huang}}{2015}]{Geng15ApJ}
{Geng} J.~J.,  {Huang} Y.~F.,  2015, \mn@doi [\apj]
  {10.1088/0004-637X/809/1/24}, \href
  {http://adsabs.harvard.edu/abs/2015ApJ...809...24G} {809, 24}

\bibitem[\protect\citeauthoryear{{Ghisellini}}{{Ghisellini}}{2017}]{Ghisellini2017MN}
{Ghisellini} G.,  2017, \mn@doi [\mnras] {10.1093/mnrasl/slw202}, \href
  {http://adsabs.harvard.edu/abs/2017MNRAS.465L..30G} {465, L30}

\bibitem[\protect\citeauthoryear{{Graham}, {Driver}, {Petrosian}, {Conselice},
  {Bershady}, {Crawford}  \& {Goto}}{{Graham}
  et~al.}{2005}]{Graham2005AJ....130.1535G}
{Graham} A.~W.,  {Driver} S.~P.,  {Petrosian} V.,  {Conselice} C.~J.,
  {Bershady} M.~A.,  {Crawford} S.~M.,   {Goto} T.,  2005, \mn@doi [\aj]
  {10.1086/444475}, \href {http://adsabs.harvard.edu/abs/2005AJ....130.1535G}
  {130, 1535}

\bibitem[\protect\citeauthoryear{Gregory}{Gregory}{2005}]{Gregory2005}
Gregory P.~C.,  2005, Bayesian Logical Data Analysis for the Physical Sciences:
  A Comparative Approach with `Mathematica' Support.
Cambridge University Press

\bibitem[\protect\citeauthoryear{{Gu}, {Dong}, {Liu}, {Ma}  \& {Wang}}{{Gu}
  et~al.}{2016}]{GDL16}
{Gu} W.-M.,  {Dong} Y.-Z.,  {Liu} T.,  {Ma} R.,   {Wang} J.,  2016, \mn@doi
  [\apjl] {10.3847/2041-8205/823/2/L28}, \href
  {http://adsabs.harvard.edu/abs/2016ApJ...823L..28G} {823, L28}

\bibitem[\protect\citeauthoryear{{Gurevich}, {Beskin}  \& {Istomin}}{{Gurevich}
  et~al.}{2006}]{GBI06PPM}
{Gurevich} A.~V.,  {Beskin} V.~S.,   {Istomin} Y.~N.,  2006, {Physics of the
  Pulsar Magnetosphere}.
Cambridge, UK: Cambridge University Press

\bibitem[\protect\citeauthoryear{{Han}, {Wang}, {Xu}  \& {Han}}{{Han}
  et~al.}{2016}]{HWX16}
{Han} J.,  {Wang} C.,  {Xu} J.,   {Han} J.-L.,  2016, \mn@doi [Research in
  Astronomy and Astrophysics] {10.1088/1674-4527/16/10/159}, \href
  {http://adsabs.harvard.edu/abs/2016RAA....16..159H} {16, 159}

\bibitem[\protect\citeauthoryear{{Hopkins} \& {Beacom}}{{Hopkins} \&
  {Beacom}}{2006}]{hb06apj}
{Hopkins} A.~M.,  {Beacom} J.~F.,  2006, \mn@doi [\apj] {10.1086/506610}, \href
  {http://adsabs.harvard.edu/abs/2006ApJ...651..142H} {651, 142}

\bibitem[\protect\citeauthoryear{{Igoshev}, {Verbunt}  \& {Cator}}{{Igoshev}
  et~al.}{2016}]{IVC16}
{Igoshev} A.,  {Verbunt} F.,   {Cator} E.,  2016, \mn@doi [\aap]
  {10.1051/0004-6361/201527471}, \href
  {http://adsabs.harvard.edu/abs/2016A%26A...591A.123I} {591, A123}

\bibitem[\protect\citeauthoryear{{Inoue}}{{Inoue}}{2004}]{Inoue04MN}
{Inoue} S.,  2004, \mn@doi [\mnras] {10.1111/j.1365-2966.2004.07359.x}, \href
  {http://adsabs.harvard.edu/abs/2004MNRAS.348..999I} {348, 999}

\bibitem[\protect\citeauthoryear{{Ioka}}{{Ioka}}{2003}]{Ioka03ApJ}
{Ioka} K.,  2003, \mn@doi [\apjl] {10.1086/380598}, \href
  {http://adsabs.harvard.edu/abs/2003ApJ...598L..79I} {598, L79}

\bibitem[\protect\citeauthoryear{{Iwazaki}}{{Iwazaki}}{2015}]{Iwazaki15PRD}
{Iwazaki} A.,  2015, \mn@doi [\prd] {10.1103/PhysRevD.91.023008}, \href
  {http://adsabs.harvard.edu/abs/2015PhRvD..91b3008I} {91, 023008}

\bibitem[\protect\citeauthoryear{{Jankowski}, {van Straten}, {Keane}, {Bailes},
  {Barr}, {Johnston}  \& {Kerr}}{{Jankowski} et~al.}{2018}]{JVK18}
{Jankowski} F.,  {van Straten} W.,  {Keane} E.~F.,  {Bailes} M.,  {Barr} E.~D.,
   {Johnston} S.,   {Kerr} M.,  2018, \mn@doi [\mnras] {10.1093/mnras/stx2476},
  \href {http://adsabs.harvard.edu/abs/2018MNRAS.473.4436J} {473, 4436}

\bibitem[\protect\citeauthoryear{{Jaynes}}{{Jaynes}}{2003}]{Jaynes03}
{Jaynes} E.~T.,  2003, {Probability Theory: The Logic of Science (Vol 1)}.
Cambridge Univ. Press, Cambridge, UK

\bibitem[\protect\citeauthoryear{{Karuppusamy}, {Stappers}  \&
  {Lee}}{{Karuppusamy} et~al.}{2012}]{KSL12}
{Karuppusamy} R.,  {Stappers} B.~W.,   {Lee} K.~J.,  2012, \mn@doi [\aap]
  {10.1051/0004-6361/201117667}, \href
  {http://adsabs.harvard.edu/abs/2012A%26A...538A...7K} {538, A7}

\bibitem[\protect\citeauthoryear{{Kashiyama}, {Ioka}  \&
  {M{\'e}sz{\'a}ros}}{{Kashiyama} et~al.}{2013}]{Kashiyama13ApJ}
{Kashiyama} K.,  {Ioka} K.,   {M{\'e}sz{\'a}ros} P.,  2013, \mn@doi [\apjl]
  {10.1088/2041-8205/776/2/L39}, \href
  {http://adsabs.harvard.edu/abs/2013ApJ...776L..39K} {776, L39}

\bibitem[\protect\citeauthoryear{{Katz}}{{Katz}}{2016}]{Katz16ApJ}
{Katz} J.~I.,  2016, \mn@doi [\apj] {10.3847/0004-637X/826/2/226}, \href
  {http://adsabs.harvard.edu/abs/2016ApJ...826..226K} {826, 226}

\bibitem[\protect\citeauthoryear{{Katz}}{{Katz}}{2017}]{Katz17MN}
{Katz} J.~I.,  2017, \mn@doi [\mnras] {10.1093/mnrasl/slx113}, \href
  {http://adsabs.harvard.edu/abs/2017MNRAS.471L..92K} {471, L92}

\bibitem[\protect\citeauthoryear{{Keane}, {Kramer}, {Lyne}, {Stappers}  \&
  {McLaughlin}}{{Keane} et~al.}{2011}]{Keane11MN}
{Keane} E.~F.,  {Kramer} M.,  {Lyne} A.~G.,  {Stappers} B.~W.,   {McLaughlin}
  M.~A.,  2011, \mn@doi [\mnras] {10.1111/j.1365-2966.2011.18917.x}, \href
  {http://adsabs.harvard.edu/abs/2011MNRAS.415.3065K} {415, 3065}

\bibitem[\protect\citeauthoryear{{Keane}, {Stappers}, {Kramer}  \&
  {Lyne}}{{Keane} et~al.}{2012}]{Keane12MN}
{Keane} E.~F.,  {Stappers} B.~W.,  {Kramer} M.,   {Lyne} A.~G.,  2012, \mn@doi
  [\mnras] {10.1111/j.1745-3933.2012.01306.x}, \href
  {http://adsabs.harvard.edu/abs/2012MNRAS.425L..71K} {425, L71}

\bibitem[\protect\citeauthoryear{{Keane} et~al.,}{{Keane}
  et~al.}{2016}]{Keane16Nat}
{Keane} E.~F.,  et~al., 2016, \mn@doi [\nat] {10.1038/nature17140}, \href
  {http://adsabs.harvard.edu/abs/2016Natur.530..453K} {530, 453}

\bibitem[\protect\citeauthoryear{{Kelly}, {Fan}  \& {Vestergaard}}{{Kelly}
  et~al.}{2008}]{KFV08}
{Kelly} B.~C.,  {Fan} X.,   {Vestergaard} M.,  2008, \mn@doi [\apj]
  {10.1086/589501}, \href {http://adsabs.harvard.edu/abs/2008ApJ...682..874K}
  {682, 874}

\bibitem[\protect\citeauthoryear{{Kennicutt}}{{Kennicutt}}{1998}]{Kenn98}
{Kennicutt} Jr. R.~C.,  1998, \mn@doi [\araa] {10.1146/annurev.astro.36.1.189},
  \href {http://adsabs.harvard.edu/abs/1998ARA%26A..36..189K} {36, 189}

\bibitem[\protect\citeauthoryear{{Kennicutt} \& {Evans}}{{Kennicutt} \&
  {Evans}}{2012}]{KE2012AR}
{Kennicutt} R.~C.,  {Evans} N.~J.,  2012, \mn@doi [\araa]
  {10.1146/annurev-astro-081811-125610}, \href
  {http://adsabs.harvard.edu/abs/2012ARA%26A..50..531K} {50, 531}

\bibitem[\protect\citeauthoryear{{Kennicutt}, {Tamblyn}  \&
  {Congdon}}{{Kennicutt} et~al.}{1994}]{KTC94ApJ}
{Kennicutt} Jr. R.~C.,  {Tamblyn} P.,   {Congdon} C.~E.,  1994, \mn@doi [\apj]
  {10.1086/174790}, \href {http://adsabs.harvard.edu/abs/1994ApJ...435...22K}
  {435, 22}

\bibitem[\protect\citeauthoryear{{Kokubo} et~al.,}{{Kokubo}
  et~al.}{2017}]{Kokubo17ApJ}
{Kokubo} M.,  et~al., 2017, \mn@doi [\apj] {10.3847/1538-4357/aa7b2d}, \href
  {http://adsabs.harvard.edu/abs/2017ApJ...844...95K} {844, 95}

\bibitem[\protect\citeauthoryear{{Landau} \& {Lifshitz}}{{Landau} \&
  {Lifshitz}}{1960}]{LandauEM}
{Landau} L.~D.,  {Lifshitz} E.~M.,  1960, {Electrodynamics of continuous
  media}.
Oxford: Pergamon Press

\bibitem[\protect\citeauthoryear{{Lentati} et~al.,}{{Lentati}
  et~al.}{2015}]{LTM16}
{Lentati} L.,  et~al., 2015, \mn@doi [\mnras] {10.1093/mnras/stv1538}, \href
  {http://adsabs.harvard.edu/abs/2015MNRAS.453.2576L} {453, 2576}

\bibitem[\protect\citeauthoryear{{Liu}, {Romero}, {Liu}  \& {Li}}{{Liu}
  et~al.}{2016}]{Liu16ApJ}
{Liu} T.,  {Romero} G.~E.,  {Liu} M.-L.,   {Li} A.,  2016, \mn@doi [\apj]
  {10.3847/0004-637X/826/1/82}, \href
  {http://adsabs.harvard.edu/abs/2016ApJ...826...82L} {826, 82}

\bibitem[\protect\citeauthoryear{{Loeb}, {Shvartzvald}  \& {Maoz}}{{Loeb}
  et~al.}{2014}]{Loeb14MN}
{Loeb} A.,  {Shvartzvald} Y.,   {Maoz} D.,  2014, \mn@doi [\mnras]
  {10.1093/mnrasl/slt177}, \href
  {http://adsabs.harvard.edu/abs/2014MNRAS.439L..46L} {439, L46}

\bibitem[\protect\citeauthoryear{{Lominadze} \& {Pataraia}}{{Lominadze} \&
  {Pataraia}}{1982}]{LP82PhST}
{Lominadze} J.~G.,  {Pataraia} A.~D.,  1982, \mn@doi [Physica Scripta Volume T]
  {10.1088/0031-8949/1982/T2A/028}, \href
  {http://adsabs.harvard.edu/abs/1982PhST....2..215L} {2, 215}

\bibitem[\protect\citeauthoryear{{Lorimer} \& {Kramer}}{{Lorimer} \&
  {Kramer}}{2012}]{LK12HPA}
{Lorimer} D.~R.,  {Kramer} M.,  2012, {Handbook of Pulsar Astronomy}.
Cambridge University Press, Cambridge, UK

\bibitem[\protect\citeauthoryear{{Lorimer}, {Bailes}, {McLaughlin}, {Narkevic}
  \& {Crawford}}{{Lorimer} et~al.}{2007}]{Lorimer07Sci}
{Lorimer} D.~R.,  {Bailes} M.,  {McLaughlin} M.~A.,  {Narkevic} D.~J.,
  {Crawford} F.,  2007, \mn@doi [Science] {10.1126/science.1147532}, \href
  {http://adsabs.harvard.edu/abs/2007Sci...318..777L} {318, 777}

\bibitem[\protect\citeauthoryear{{Lu} \& {Kumar}}{{Lu} \&
  {Kumar}}{2018}]{Lu18MN}
{Lu} W.,  {Kumar} P.,  2018, \mn@doi [\mnras] {10.1093/mnras/sty716}, \href
  {http://adsabs.harvard.edu/abs/2018MNRAS.477.2470L} {477, 2470}

\bibitem[\protect\citeauthoryear{{Lynden-Bell}}{{Lynden-Bell}}{1971}]{LB71}
{Lynden-Bell} D.,  1971, \mn@doi [\mnras] {10.1093/mnras/155.1.95}, \href
  {http://adsabs.harvard.edu/abs/1971MNRAS.155...95L} {155, 95}

\bibitem[\protect\citeauthoryear{{Lyubarsky}}{{Lyubarsky}}{2014}]{Lyubarsky14MN}
{Lyubarsky} Y.,  2014, \mn@doi [\mnras] {10.1093/mnrasl/slu046}, \href
  {http://adsabs.harvard.edu/abs/2014MNRAS.442L...9L} {442, L9}

\bibitem[\protect\citeauthoryear{{Macquart} et~al.,}{{Macquart}
  et~al.}{2010}]{Macquart10PASA}
{Macquart} J.-P.,  et~al., 2010, \mn@doi [\pasa] {10.1071/AS09082}, \href
  {http://adsabs.harvard.edu/abs/2010PASA...27..272M} {27, 272}

\bibitem[\protect\citeauthoryear{{Macquart} et~al.,}{{Macquart}
  et~al.}{2015}]{Macquart15SKA}
{Macquart} J.~P.,  et~al., 2015, Advancing Astrophysics with the Square
  Kilometre Array (AASKA14), \href
  {http://adsabs.harvard.edu/abs/2015aska.confE..55M} {p.~55}

\bibitem[\protect\citeauthoryear{{Madau} \& {Dickinson}}{{Madau} \&
  {Dickinson}}{2014}]{MD14}
{Madau} P.,  {Dickinson} M.,  2014, \mn@doi [\araa]
  {10.1146/annurev-astro-081811-125615}, \href
  {http://adsabs.harvard.edu/abs/2014ARA%26A..52..415M} {52, 415}

\bibitem[\protect\citeauthoryear{{Madau}, {Pozzetti}  \& {Dickinson}}{{Madau}
  et~al.}{1998}]{MPD98ApJ}
{Madau} P.,  {Pozzetti} L.,   {Dickinson} M.,  1998, \mn@doi [\apj]
  {10.1086/305523}, \href {http://adsabs.harvard.edu/abs/1998ApJ...498..106M}
  {498, 106}

\bibitem[\protect\citeauthoryear{{Manchester}, {Fan}, {Lyne}, {Kaspi}  \&
  {Crawford}}{{Manchester} et~al.}{2006}]{MFL06}
{Manchester} R.~N.,  {Fan} G.,  {Lyne} A.~G.,  {Kaspi} V.~M.,   {Crawford} F.,
  2006, \mn@doi [\apj] {10.1086/505461}, \href
  {http://adsabs.harvard.edu/abs/2006ApJ...649..235M} {649, 235}

\bibitem[\protect\citeauthoryear{{Marshall}, {Tananbaum}, {Avni}  \&
  {Zamorani}}{{Marshall} et~al.}{1983}]{MTA83}
{Marshall} H.~L.,  {Tananbaum} H.,  {Avni} Y.,   {Zamorani} G.,  1983, \mn@doi
  [\apj] {10.1086/161016}, \href
  {http://adsabs.harvard.edu/abs/1983ApJ...269...35M} {269, 35}

\bibitem[\protect\citeauthoryear{{Masui} \& {Sigurdson}}{{Masui} \&
  {Sigurdson}}{2015}]{Masui15PRL}
{Masui} K.~W.,  {Sigurdson} K.,  2015, \mn@doi [Physical Review Letters]
  {10.1103/PhysRevLett.115.121301}, \href
  {http://adsabs.harvard.edu/abs/2015PhRvL.115l1301M} {115, 121301}

\bibitem[\protect\citeauthoryear{{Masui} et~al.,}{{Masui}
  et~al.}{2015}]{Masui15Nat}
{Masui} K.,  et~al., 2015, \mn@doi [\nat] {10.1038/nature15769}, \href
  {http://adsabs.harvard.edu/abs/2015Natur.528..523M} {528, 523}

\bibitem[\protect\citeauthoryear{{McQuinn}}{{McQuinn}}{2014}]{McQuinn14ApJ}
{McQuinn} M.,  2014, \mn@doi [\apjl] {10.1088/2041-8205/780/2/L33}, \href
  {http://adsabs.harvard.edu/abs/2014ApJ...780L..33M} {780, L33}

\bibitem[\protect\citeauthoryear{{McQuinn}, {Lidz}, {Zaldarriaga}, {Hernquist},
  {Hopkins}, {Dutta}  \& {Faucher-Gigu{\`e}re}}{{McQuinn}
  et~al.}{2009}]{McQuinn09ApJ}
{McQuinn} M.,  {Lidz} A.,  {Zaldarriaga} M.,  {Hernquist} L.,  {Hopkins} P.~F.,
   {Dutta} S.,   {Faucher-Gigu{\`e}re} C.-A.,  2009, \mn@doi [\apj]
  {10.1088/0004-637X/694/2/842}, \href
  {http://adsabs.harvard.edu/abs/2009ApJ...694..842M} {694, 842}

\bibitem[\protect\citeauthoryear{{Metzger}, {Berger}  \& {Margalit}}{{Metzger}
  et~al.}{2017}]{Metzger17ApJ}
{Metzger} B.~D.,  {Berger} E.,   {Margalit} B.,  2017, \mn@doi [\apj]
  {10.3847/1538-4357/aa633d}, \href
  {http://adsabs.harvard.edu/abs/2017ApJ...841...14M} {841, 14}

\bibitem[\protect\citeauthoryear{{Nakamura}, {Fukugita}, {Yasuda}, {Loveday},
  {Brinkmann}, {Schneider}, {Shimasaku}  \& {SubbaRao}}{{Nakamura}
  et~al.}{2003}]{Nakamura03AJ}
{Nakamura} O.,  {Fukugita} M.,  {Yasuda} N.,  {Loveday} J.,  {Brinkmann} J.,
  {Schneider} D.~P.,  {Shimasaku} K.,   {SubbaRao} M.,  2003, \mn@doi [\aj]
  {10.1086/368135}, \href {http://adsabs.harvard.edu/abs/2003AJ....125.1682N}
  {125, 1682}

\bibitem[\protect\citeauthoryear{{Nakamura}, {Fukugita}, {Brinkmann}  \&
  {Schneider}}{{Nakamura} et~al.}{2004}]{Nakamura04AJ}
{Nakamura} O.,  {Fukugita} M.,  {Brinkmann} J.,   {Schneider} D.~P.,  2004,
  \mn@doi [\aj] {10.1086/386350}, \href
  {http://adsabs.harvard.edu/abs/2004AJ....127.2511N} {127, 2511}

\bibitem[\protect\citeauthoryear{{Nan} et~al.,}{{Nan} et~al.}{2011}]{Nan11}
{Nan} R.,  et~al., 2011, \mn@doi [International Journal of Modern Physics D]
  {10.1142/S0218271811019335}, \href
  {http://adsabs.harvard.edu/abs/2011IJMPD..20..989N} {20, 989}

\bibitem[\protect\citeauthoryear{{Ng} et~al.,}{{Ng} et~al.}{2017}]{Ng17CHIME}
{Ng} C.,  et~al., 2017, preprint, \href
  {http://adsabs.harvard.edu/abs/2017arXiv170204728N} {} (\mn@eprint {arXiv}
  {1702.04728})

\bibitem[\protect\citeauthoryear{{Niino}}{{Niino}}{2018}]{Niino18ApJ}
{Niino} Y.,  2018, \mn@doi [\apj] {10.3847/1538-4357/aab9a9}, \href
  {http://adsabs.harvard.edu/abs/2018ApJ...858....4N} {858, 4}

\bibitem[\protect\citeauthoryear{{Oslowski} et~al.,}{{Oslowski}
  et~al.}{2018a}]{Oslowski18ATel1}
{Oslowski} S.,  et~al., 2018a, The Astronomer's Telegram, \href
  {http://adsabs.harvard.edu/abs/2018ATel11385....1O} {11385}

\bibitem[\protect\citeauthoryear{{Oslowski} et~al.,}{{Oslowski}
  et~al.}{2018b}]{Oslowski18ATel2}
{Oslowski} S.,  et~al., 2018b, The Astronomer's Telegram, \href
  {http://adsabs.harvard.edu/abs/2018ATel11396....1O} {11396}

\bibitem[\protect\citeauthoryear{{Pen} \& {Connor}}{{Pen} \&
  {Connor}}{2015}]{Pen15ApJ}
{Pen} U.-L.,  {Connor} L.,  2015, \mn@doi [\apj] {10.1088/0004-637X/807/2/179},
  \href {http://adsabs.harvard.edu/abs/2015ApJ...807..179P} {807, 179}

\bibitem[\protect\citeauthoryear{{Pescalli} et~al.,}{{Pescalli}
  et~al.}{2016}]{PGS16}
{Pescalli} A.,  et~al., 2016, \mn@doi [\aap] {10.1051/0004-6361/201526760},
  \href {http://adsabs.harvard.edu/abs/2016A%26A...587A..40P} {587, A40}

\bibitem[\protect\citeauthoryear{{Petroff} et~al.,}{{Petroff}
  et~al.}{2015}]{Petroff15MN}
{Petroff} E.,  et~al., 2015, \mn@doi [\mnras] {10.1093/mnras/stu2419}, \href
  {http://adsabs.harvard.edu/abs/2015MNRAS.447..246P} {447, 246}

\bibitem[\protect\citeauthoryear{{Petroff} et~al.,}{{Petroff}
  et~al.}{2016}]{Petroff16PASA}
{Petroff} E.,  et~al., 2016, \mn@doi [\pasa] {10.1017/pasa.2016.35}, \href
  {http://adsabs.harvard.edu/abs/2016PASA...33...45P} {33, e045}

\bibitem[\protect\citeauthoryear{{Petroff} et~al.,}{{Petroff}
  et~al.}{2017}]{Petroff17MN}
{Petroff} E.,  et~al., 2017, \mn@doi [\mnras] {10.1093/mnras/stx1098}, \href
  {http://adsabs.harvard.edu/abs/2017MNRAS.469.4465P} {469, 4465}

\bibitem[\protect\citeauthoryear{{Petrosian}}{{Petrosian}}{1976}]{Petrosian76ApJ}
{Petrosian} V.,  1976, \mn@doi [\apjl] {10.1086/182253}, \href
  {http://adsabs.harvard.edu/abs/1976ApJ...209L...1P} {209, L1}

\bibitem[\protect\citeauthoryear{{Planck Collaboration} et~al.,}{{Planck
  Collaboration} et~al.}{2016}]{Planck16A&A}
{Planck Collaboration} et~al., 2016, \mn@doi [\aap]
  {10.1051/0004-6361/201525830}, \href
  {http://adsabs.harvard.edu/abs/2016A%26A...594A..13P} {594, A13}

\bibitem[\protect\citeauthoryear{{Popov} \& {Postnov}}{{Popov} \&
  {Postnov}}{2010}]{PP10}
{Popov} S.~B.,  {Postnov} K.~A.,  2010, in {Harutyunian} H.~A.,  {Mickaelian}
  A.~M.,   {Terzian} Y.,  eds, Evolution of Cosmic Objects through their
  Physical Activity. pp 129--132 (\mn@eprint {arXiv} {0710.2006})

\bibitem[\protect\citeauthoryear{{Popov} \& {Postnov}}{{Popov} \&
  {Postnov}}{2013}]{PP13arXiv}
{Popov} S.~B.,  {Postnov} K.~A.,  2013, preprint, \href
  {http://adsabs.harvard.edu/abs/2013arXiv1307.4924P} {} (\mn@eprint {arXiv}
  {1307.4924})

\bibitem[\protect\citeauthoryear{{Price} et~al.,}{{Price}
  et~al.}{2018}]{Price18ATel}
{Price} D.~C.,  et~al., 2018, The Astronomer's Telegram, \href
  {http://adsabs.harvard.edu/abs/2018ATel11376....1P} {11376}

\bibitem[\protect\citeauthoryear{{Ravi}, {Shannon}  \& {Jameson}}{{Ravi}
  et~al.}{2015}]{Ravi15ApJ}
{Ravi} V.,  {Shannon} R.~M.,   {Jameson} A.,  2015, \mn@doi [\apjl]
  {10.1088/2041-8205/799/1/L5}, \href
  {http://adsabs.harvard.edu/abs/2015ApJ...799L...5R} {799, L5}

\bibitem[\protect\citeauthoryear{{Ravi} et~al.,}{{Ravi}
  et~al.}{2016}]{Ravi16Sci}
{Ravi} V.,  et~al., 2016, \mn@doi [Science] {10.1126/science.aaf6807}, \href
  {http://adsabs.harvard.edu/abs/2016Sci...354.1249R} {354, 1249}

\bibitem[\protect\citeauthoryear{{Rees}}{{Rees}}{1977}]{Rees77Nat}
{Rees} M.~J.,  1977, \mn@doi [\nat] {10.1038/266333a0}, \href
  {http://adsabs.harvard.edu/abs/1977Natur.266..333R} {266, 333}

\bibitem[\protect\citeauthoryear{{Reynolds}}{{Reynolds}}{1977}]{Reynolds77ApJ}
{Reynolds} R.~J.,  1977, \mn@doi [\apj] {10.1086/155484}, \href
  {http://adsabs.harvard.edu/abs/1977ApJ...216..433R} {216, 433}

\bibitem[\protect\citeauthoryear{{Romero}, {del Valle}  \& {Vieyro}}{{Romero}
  et~al.}{2016}]{Romero16PRD}
{Romero} G.~E.,  {del Valle} M.~V.,   {Vieyro} F.~L.,  2016, \mn@doi [\prd]
  {10.1103/PhysRevD.93.023001}, \href
  {http://adsabs.harvard.edu/abs/2016PhRvD..93b3001R} {93, 023001}

\bibitem[\protect\citeauthoryear{{Rybicki} \& {Lightman}}{{Rybicki} \&
  {Lightman}}{1986}]{RL86}
{Rybicki} G.~B.,  {Lightman} A.~P.,  1986, {Radiative Processes in
  Astrophysics}.
Wiley-VCH

\bibitem[\protect\citeauthoryear{{Schechter}}{{Schechter}}{1976}]{Schechter76ApJ}
{Schechter} P.,  1976, \mn@doi [\apj] {10.1086/154079}, \href
  {http://adsabs.harvard.edu/abs/1976ApJ...203..297S} {203, 297}

\bibitem[\protect\citeauthoryear{{Scholz} et~al.,}{{Scholz}
  et~al.}{2016}]{Scholz16ApJ}
{Scholz} P.,  et~al., 2016, \mn@doi [\apj] {10.3847/1538-4357/833/2/177}, \href
  {http://adsabs.harvard.edu/abs/2016ApJ...833..177S} {833, 177}

\bibitem[\protect\citeauthoryear{{Sembach} et~al.,}{{Sembach}
  et~al.}{2003}]{SWS03}
{Sembach} K.~R.,  et~al., 2003, \mn@doi [\apjs] {10.1086/346231}, \href
  {http://adsabs.harvard.edu/abs/2003ApJS..146..165S} {146, 165}

\bibitem[\protect\citeauthoryear{{Shand}, {Ouyed}, {Koning}  \&
  {Ouyed}}{{Shand} et~al.}{2016}]{Shand16RAA}
{Shand} Z.,  {Ouyed} A.,  {Koning} N.,   {Ouyed} R.,  2016, \mn@doi [Research
  in Astronomy and Astrophysics] {10.1088/1674-4527/16/5/080}, \href
  {http://adsabs.harvard.edu/abs/2016RAA....16...80S} {16, 80}

\bibitem[\protect\citeauthoryear{{Shannon} et~al.,}{{Shannon}
  et~al.}{2017}]{Shannon17ATel}
{Shannon} R.~M.,  et~al., 2017, The Astronomer's Telegram, \href
  {http://adsabs.harvard.edu/abs/2017ATel11046....1S} {11046}

\bibitem[\protect\citeauthoryear{{Shao} \& {Zhang}}{{Shao} \&
  {Zhang}}{2017}]{Shao17PRD}
{Shao} L.,  {Zhang} B.,  2017, \mn@doi [\prd] {10.1103/PhysRevD.95.123010},
  \href {http://adsabs.harvard.edu/abs/2017PhRvD..95l3010S} {95, 123010}

\bibitem[\protect\citeauthoryear{{Shen}, {Mo}, {White}, {Blanton}, {Kauffmann},
  {Voges}, {Brinkmann}  \& {Csabai}}{{Shen} et~al.}{2003}]{Shen03MN}
{Shen} S.,  {Mo} H.~J.,  {White} S.~D.~M.,  {Blanton} M.~R.,  {Kauffmann} G.,
  {Voges} W.,  {Brinkmann} J.,   {Csabai} I.,  2003, \mn@doi [\mnras]
  {10.1046/j.1365-8711.2003.06740.x}, \href
  {http://adsabs.harvard.edu/abs/2003MNRAS.343..978S} {343, 978}

\bibitem[\protect\citeauthoryear{{Skilling}}{{Skilling}}{2004}]{Skilling04}
{Skilling} J.,  2004, in {Fischer} R.,  {Preuss} R.,   {Toussaint} U.~V.,  eds,
   American Institute of Physics Conference Series Vol. 735, American Institute
  of Physics Conference Series. pp 395--405, \mn@doi{10.1063/1.1835238}

\bibitem[\protect\citeauthoryear{{Spitler} et~al.,}{{Spitler}
  et~al.}{2014}]{Spitler14ApJ}
{Spitler} L.~G.,  et~al., 2014, \mn@doi [\apj] {10.1088/0004-637X/790/2/101},
  \href {http://adsabs.harvard.edu/abs/2014ApJ...790..101S} {790, 101}

\bibitem[\protect\citeauthoryear{{Spitler} et~al.,}{{Spitler}
  et~al.}{2016}]{Spitler16Nat}
{Spitler} L.~G.,  et~al., 2016, \mn@doi [\nat] {10.1038/nature17168}, \href
  {http://adsabs.harvard.edu/abs/2016Natur.531..202S} {531, 202}

\bibitem[\protect\citeauthoryear{{Sun}, {Zhang}  \& {Li}}{{Sun}
  et~al.}{2015}]{SZL15}
{Sun} H.,  {Zhang} B.,   {Li} Z.,  2015, \mn@doi [\apj]
  {10.1088/0004-637X/812/1/33}, \href
  {http://adsabs.harvard.edu/abs/2015ApJ...812...33S} {812, 33}

\bibitem[\protect\citeauthoryear{{Tendulkar} et~al.,}{{Tendulkar}
  et~al.}{2017}]{Tendulkar17ApJ}
{Tendulkar} S.~P.,  et~al., 2017, \mn@doi [\apjl] {10.3847/2041-8213/834/2/L7},
  \href {http://adsabs.harvard.edu/abs/2017ApJ...834L...7T} {834, L7}

\bibitem[\protect\citeauthoryear{{Thornton}}{{Thornton}}{2013}]{Thornton13PhDT}
{Thornton} D.,  2013, PhD thesis, University of Manchester

\bibitem[\protect\citeauthoryear{{Thornton} et~al.,}{{Thornton}
  et~al.}{2013}]{Thornton13Sci}
{Thornton} D.,  et~al., 2013, \mn@doi [Science] {10.1126/science.1236789},
  \href {http://adsabs.harvard.edu/abs/2013Sci...341...53T} {341, 53}

\bibitem[\protect\citeauthoryear{{Tingay} \& {Kaplan}}{{Tingay} \&
  {Kaplan}}{2016}]{Tingay16ApJ}
{Tingay} S.~J.,  {Kaplan} D.~L.,  2016, \mn@doi [\apjl]
  {10.3847/2041-8205/820/2/L31}, \href
  {http://adsabs.harvard.edu/abs/2016ApJ...820L..31T} {820, L31}

\bibitem[\protect\citeauthoryear{{Totani}}{{Totani}}{2013}]{Totani2013PASJ}
{Totani} T.,  2013, \mn@doi [\pasj] {10.1093/pasj/65.5.L12}, \href
  {http://adsabs.harvard.edu/abs/2013PASJ...65L..12T} {65, L12}

\bibitem[\protect\citeauthoryear{{Verbiest}, {Lorimer}  \&
  {McLaughlin}}{{Verbiest} et~al.}{2010}]{VLM10MN}
{Verbiest} J.~P.~W.,  {Lorimer} D.~R.,   {McLaughlin} M.~A.,  2010, \mn@doi
  [\mnras] {10.1111/j.1365-2966.2010.16488.x}, \href
  {http://adsabs.harvard.edu/abs/2010MNRAS.405..564V} {405, 564}

\bibitem[\protect\citeauthoryear{{Verbiest}, {Weisberg}, {Chael}, {Lee}  \&
  {Lorimer}}{{Verbiest} et~al.}{2012}]{Verbiest12ApJ}
{Verbiest} J.~P.~W.,  {Weisberg} J.~M.,  {Chael} A.~A.,  {Lee} K.~J.,
  {Lorimer} D.~R.,  2012, \mn@doi [\apj] {10.1088/0004-637X/755/1/39}, \href
  {http://adsabs.harvard.edu/abs/2012ApJ...755...39V} {755, 39}

\bibitem[\protect\citeauthoryear{{Vieyro}, {Romero}, {Bosch-Ramon}, {Marcote}
  \& {del Valle}}{{Vieyro} et~al.}{2017}]{Vieyro17A&A}
{Vieyro} F.~L.,  {Romero} G.~E.,  {Bosch-Ramon} V.,  {Marcote} B.,   {del
  Valle} M.~V.,  2017, \mn@doi [\aap] {10.1051/0004-6361/201730556}, \href
  {http://adsabs.harvard.edu/abs/2017A%26A...602A..64V} {602, A64}

\bibitem[\protect\citeauthoryear{{Wang}}{{Wang}}{2017}]{Wang17}
{Wang} N.,  2017, \mn@doi [Scientia Sinica Physica, Mechanica \& Astronomica]
  {10.1360/SSPMA2017-00082}, \href
  {http://adsabs.harvard.edu/abs/2017SSPMA..47e9501W} {47, 059501}

\bibitem[\protect\citeauthoryear{{Wang}, {Yang}, {Wu}, {Dai}  \& {Wang}}{{Wang}
  et~al.}{2016}]{Wang16ApJ}
{Wang} J.-S.,  {Yang} Y.-P.,  {Wu} X.-F.,  {Dai} Z.-G.,   {Wang} F.-Y.,  2016,
  \mn@doi [\apjl] {10.3847/2041-8205/822/1/L7}, \href
  {http://adsabs.harvard.edu/abs/2016ApJ...822L...7W} {822, L7}

\bibitem[\protect\citeauthoryear{{Wang}, {Luo}, {Yue}, {Chen}, {Lee}  \&
  {Xu}}{{Wang} et~al.}{2018}]{Wang18ApJ}
{Wang} W.,  {Luo} R.,  {Yue} H.,  {Chen} X.,  {Lee} K.,   {Xu} R.,  2018,
  \mn@doi [\apj] {10.3847/1538-4357/aaa025}, \href
  {http://adsabs.harvard.edu/abs/2018ApJ...852..140W} {852, 140}

\bibitem[\protect\citeauthoryear{{Wei}, {Gao}, {Wu}  \&
  {M{\'e}sz{\'a}ros}}{{Wei} et~al.}{2015}]{Wei15PRL}
{Wei} J.-J.,  {Gao} H.,  {Wu} X.-F.,   {M{\'e}sz{\'a}ros} P.,  2015, \mn@doi
  [Physical Review Letters] {10.1103/PhysRevLett.115.261101}, \href
  {http://adsabs.harvard.edu/abs/2015PhRvL.115z1101W} {115, 261101}

\bibitem[\protect\citeauthoryear{{Willmer}}{{Willmer}}{1997}]{Will97}
{Willmer} C.~N.~A.,  1997, \mn@doi [\aj] {10.1086/118522}, \href
  {http://adsabs.harvard.edu/abs/1997AJ....114..898W} {114, 898}

\bibitem[\protect\citeauthoryear{{Wu} et~al.,}{{Wu} et~al.}{2016}]{Wu16ApJ}
{Wu} X.-F.,  et~al., 2016, \mn@doi [\apjl] {10.3847/2041-8205/822/1/L15}, \href
  {http://adsabs.harvard.edu/abs/2016ApJ...822L..15W} {822, L15}

\bibitem[\protect\citeauthoryear{{Xu} \& {Han}}{{Xu} \& {Han}}{2015}]{XH15RAA}
{Xu} J.,  {Han} J.~L.,  2015, \mn@doi [Research in Astronomy and Astrophysics]
  {10.1088/1674-4527/15/10/002}, \href
  {http://adsabs.harvard.edu/abs/2015RAA....15.1629X} {15, 1629}

\bibitem[\protect\citeauthoryear{{Yang}, {Luo}, {Li}  \& {Zhang}}{{Yang}
  et~al.}{2017}]{Yang17ApJ}
{Yang} Y.-P.,  {Luo} R.,  {Li} Z.,   {Zhang} B.,  2017, \mn@doi [\apjl]
  {10.3847/2041-8213/aa6c2e}, \href
  {http://adsabs.harvard.edu/abs/2017ApJ...839L..25Y} {839, L25}

\bibitem[\protect\citeauthoryear{{Yao}, {Manchester}  \& {Wang}}{{Yao}
  et~al.}{2017}]{YMW16}
{Yao} J.~M.,  {Manchester} R.~N.,   {Wang} N.,  2017, \mn@doi [\apj]
  {10.3847/1538-4357/835/1/29}, \href
  {http://adsabs.harvard.edu/abs/2017ApJ...835...29Y} {835, 29}

\bibitem[\protect\citeauthoryear{{Young}}{{Young}}{1976}]{Young76AJ}
{Young} P.~J.,  1976, \mn@doi [\aj] {10.1086/111959}, \href
  {http://adsabs.harvard.edu/abs/1976AJ.....81..807Y} {81, 807}

\bibitem[\protect\citeauthoryear{{Yu}}{{Yu}}{2014}]{Yu14ApJ}
{Yu} Y.-W.,  2014, \mn@doi [\apj] {10.1088/0004-637X/796/2/93}, \href
  {http://adsabs.harvard.edu/abs/2014ApJ...796...93Y} {796, 93}

\bibitem[\protect\citeauthoryear{{Yu}, {Cheng}, {Shiu}  \& {Tye}}{{Yu}
  et~al.}{2014}]{Yu14JCAP}
{Yu} Y.-W.,  {Cheng} K.-S.,  {Shiu} G.,   {Tye} H.,  2014, \mn@doi [\jcap]
  {10.1088/1475-7516/2014/11/040}, \href
  {http://adsabs.harvard.edu/abs/2014JCAP...11..040Y} {11, 40}

\bibitem[\protect\citeauthoryear{{Zeilinger}, {M{\o}ller}  \&
  {Stiavelli}}{{Zeilinger} et~al.}{1993}]{ZMS93MN}
{Zeilinger} W.~W.,  {M{\o}ller} P.,   {Stiavelli} M.,  1993, \mn@doi [\mnras]
  {10.1093/mnras/261.1.175}, \href
  {http://adsabs.harvard.edu/abs/1993MNRAS.261..175Z} {261, 175}

\bibitem[\protect\citeauthoryear{{Zhang}}{{Zhang}}{2014}]{Zhang14ApJ}
{Zhang} B.,  2014, \mn@doi [\apjl] {10.1088/2041-8205/780/2/L21}, \href
  {http://adsabs.harvard.edu/abs/2014ApJ...780L..21Z} {780, L21}

\bibitem[\protect\citeauthoryear{{Zhang}}{{Zhang}}{2016a}]{Zhang16arXiv}
{Zhang} S.-N.,  2016a, preprint, \href
  {http://adsabs.harvard.edu/abs/2016arXiv160104558Z} {} (\mn@eprint {arXiv}
  {1601.04558})

\bibitem[\protect\citeauthoryear{{Zhang}}{{Zhang}}{2016b}]{Zhang16ApJL}
{Zhang} B.,  2016b, \mn@doi [\apjl] {10.3847/2041-8205/827/2/L31}, \href
  {http://adsabs.harvard.edu/abs/2016ApJ...827L..31Z} {827, L31}

\bibitem[\protect\citeauthoryear{{Zhang}}{{Zhang}}{2017}]{Zhang17ApJ}
{Zhang} B.,  2017, \mn@doi [\apjl] {10.3847/2041-8213/aa5ded}, \href
  {http://adsabs.harvard.edu/abs/2017ApJ...836L..32Z} {836, L32}

\bibitem[\protect\citeauthoryear{{Zhang}}{{Zhang}}{2018a}]{Zhang18arXiv}
{Zhang} B.,  2018a, preprint, \href
  {http://adsabs.harvard.edu/abs/2018arXiv180805277Z} {} (\mn@eprint {arXiv}
  {1808.05277})

\bibitem[\protect\citeauthoryear{{Zhang}}{{Zhang}}{2018b}]{Zhang18ApJ}
{Zhang} B.,  2018b, \mn@doi [\apjl] {10.3847/2041-8213/aaadba}, \href
  {http://adsabs.harvard.edu/abs/2018ApJ...854L..21Z} {854, L21}

\bibitem[\protect\citeauthoryear{{Zheng}, {Ofek}, {Kulkarni}, {Neill}  \&
  {Juric}}{{Zheng} et~al.}{2014}]{Zheng14ApJ}
{Zheng} Z.,  {Ofek} E.~O.,  {Kulkarni} S.~R.,  {Neill} J.~D.,   {Juric} M.,
  2014, \mn@doi [\apj] {10.1088/0004-637X/797/1/71}, \href
  {http://adsabs.harvard.edu/abs/2014ApJ...797...71Z} {797, 71}

\bibitem[\protect\citeauthoryear{{Zhou}, {Li}, {Wang}, {Fan}  \& {Wei}}{{Zhou}
  et~al.}{2014}]{Zhou14PRD}
{Zhou} B.,  {Li} X.,  {Wang} T.,  {Fan} Y.-Z.,   {Wei} D.-M.,  2014, \mn@doi
  [\prd] {10.1103/PhysRevD.89.107303}, \href
  {http://adsabs.harvard.edu/abs/2014PhRvD..89j7303Z} {89, 107303}

\makeatother
\end{thebibliography}


\appendix
\section{Notations used in the current paper}
All the notations used in this paper are listed in Table A1.

\begin{table*}
\centering
\caption{Notations used in the current paper sorted alphabetically.}
\begin{threeparttable}
\begin{tabular}{cl}
\hline
\hline
Notation	& Comments \\
\hline
$a(z)$ & Scale factor in Cosmology \\
$\alpha$ & Power-law index of FRB luminosity function \\
$\alpha_\mathrm{e}$ & Power-law index of galaxy electron density profile \\
$\rm BW$ & Bandwidth of the data, in units of MHz\\
$c$ & Speed of light in a vacuum in units of cm~s$^{-1}$\\
$\chi(z)$ & Cosmological ionisation fraction as function of redshift $z$ \\
$\Delta \nu_0=1\, {\rm GHz}$ & Reference spectrum width of FRB\\
$\Delta t$ & Time delay in units of ms\\
$\DM$ & Dispersion measure, in units of $\cmpc$\\
$\DMe$ & Extragalactic dispersion measure, i.e. $\DMe=\DM-\DMm$ \\
$\DMi$ & Dispersion measure contribution of IGM\\
$\DMd$ & Dispersion measure contribution from dark matter halo of the Milky Way \\
$\DMh$ & Dispersion measure contribution of FRB host galaxy\\
$\DMho$ & Normalised dispersion measure contribution of FRB host galaxy 
at redshift of 0 using star formation history\\
$\DMm$ & Dispersion measure contribution of the Milky way\\
$\DMs$ & Dispersion measure contribution of the local source \\
EM & Emission measure, in units of ${\rm cm}^{-6}\, {\rm pc}$\\
$\mathcal{E}$ & Energy, in units of erg \\
$E(z)$ & Logarithmic time derivative of scale factor \\
$\phi^*$ & Normalisation factor of luminosity function\\
$\epsilon$ & beam resonance of radio telescope \\
$\fd$ & Distribution function of $\DMh$ \\
$f_{\rm IGM}$ & Baryon mass fraction in the IGM \\
$f_{\rm s}$ & Distribution function of $\DMs$ \\
$F$ & Specific fluence, the total energy density of the burst, i.e. the time integrated flux density per unit frequency \\
$\phi(L)$ & Luminosity function\\
$g(z)$ & Electron density per baryon as a function of cosmological redshift $z$ \\
$G$ & Gain of radio telescope, in units of ${\rm K\, Jy^{-1}}$\\
$h$ & Dimensionless Hubble parameter, normalised by $100\,{\rm km\,s^{-1} Mpc^{-1}}$\\
$H^*$ & Scale height of disk galaxy \\
$H_0$ & Hubble constant with $H_{0}=67\, {\rm km\,s^{-1} Mpc^{-1}}$\\
$I$ & Intensity, in units of $\rm erg\, s^{-1}\, sr^{-1}\, cm^{-2}$ \\
$I(\DMe, z)$ & Marginalised integral for $\DMs$ \\
$I(\log L)$ & Marginalised integral for beam response $\epsilon$\\
$l$ & integration path length, in units of pc \\
$L$ & Luminosity, in units of $\ergs$ \\
$L^*$ & Upper cut-off luminosity \\
$L_0$ & Lower cut-off luminosity \\
$\Lambda$ & Likelihood function \\
$n_{\rm e}$ & Electron density, in units of ${\rm cm}^{-3}$ \\
$\nu$ & Observing frequency, in units of GHz \\
$n^*$ & Galaxy density in comoving volume, that is $n^*=\int \phi(L)\,\D L$ \\
$N_{\rm f}$ & Normalisation factor for likelihood function \\
$M$ & Absolute stellar magnitude\\
$\Omega_{\rm b}$ & Dimensionless baryon fraction of Universe. Assumed to be 0.048. \\
$\Omega_{\rm \Lambda}$ & Dimensionless cosmological constant. Assumed to be 0.69. \\
$\Omega_{\rm m}$ & Dimensionless matter fraction of Universe. Assumed to be 0.31. \\
$\varpi$ & Radius from the z-axis in the cylindrical coordinate.\\
$\varpi^*$ & Scale radius of the disk galaxy\\
$r$ & Comoving distance \\
$r_{\rm L}$ & Luminosity distance, in terms of comoving distance $r_{\rm L}=(1+z)r$ \\
$R$ & Radius of galaxy, in units of kpc\\
$R_{\rm e}$ & Effective radius of galaxy, in units of kpc \\
$R_{50}$ & Petrosian radius of galaxy, the radius enclosing 50 percent of petrosian flux \\
$\rho$ & Stellar density in units of ${\rm pc}^{-3}$ \\
$\rm SEFD$ & System equivalent flux density, in units of $\rm Jy$\\
${\rm SFR}(z)$ & Star formation history as function of redshift in units of $\rm M_{\rm \sun} yr^{-1}$ \\
$S_{\rm peak}$ & Peak flux density of FRB, in units of Jy \\
$T$ & Temperature of ionised gas \\
$T_{\rm morph}$ & Morphological index of galaxies \\
$T_{\rm sys}$ & System temperature of radio receiver, in units of K \\
$\VEC{\Theta}$ & General notation for parameters \\
$\theta$ & Angular distance between FRB and beam centre\\
$\theta_b$ & Beam size of radio telescope\\
$w$ & FRB duration, in units of ms \\
$\VEC{X}$ & General notation for data \\
$z$ & Cosmological redshift \\
$Z$ & Vertical distances of FRB to the galaxy disk plane\\
\hline \end{tabular} \label{tab:not} \end{threeparttable} \end{table*}

\section{Data table}
\label{app:dattab}

The data in the FRB catalogue \citep{Petroff16PASA} is given in \TAB{tab:frbs}.  
The columns noted as the \emph{Observed parameters} are from the FRB catalogue, 
while the \emph{Infered parameters} are computed using the methods in this 
paper. The inferred parameters are for reference purposes. They are not used in 
our computation for luminosity functions, as they are \emph{not} needed in the 
likelihood function \EQ{eq:likfun}. The details on how to 
calculat redshift, luminosity and energy of each FRB are 
presented in \APP{app:probl}.

\begin{landscape}
\begin{table}
\caption{The parameters of known FRBs}
\begin{threeparttable}
\begin{tabular}{ccccccccccccc}
\hline
\hline
& \multicolumn{4}{c}{Observed parameters} & \multicolumn{6}{c}{Infered 
parameters} \\
\cmidrule(lr){2-5} \cmidrule(lr){6-11} FRB & $S_\mathrm{peak}$\tnote{(a)} & 
$w$\tnote{(b)} & $F$ \tnote{(c)} & DM \tnote{(d)} & $\DMe$\tnote{(e)} & 
$\DMe$\tnote{(f)}  & $z_{\rm max}$\tnote{(g)} & $\hat{z}$\tnote{(h)} & 
$\log \hat{L}_{\rm iso}$\tnote{(i)}& $ \log \hat{\mathcal{E}}_{\rm 
iso}$\tnote{(j)}& Survey & Reference \\
& (Jy) & (ms) & (Jy ms) & $(\cmpc)$& $(\cmpc)$ & $(\cmpc)$ & & & 
($\ergs$)& (erg) & &\\
\hline
010125 & 0.30 & $9.40^{+0.20}_{-0.20}$ & 2.82 & 790(3) & 680 & 714.09 &                                                 $0.80$ & $0.765^{+0.005}_{-0.091}$ & $43.00^{+0.29}_{-0.34}$ & $40.75^{+0.24}_{-0.33}$     & Parkes I & [1] \\
010621 & 0.41 & 7.00 & 2.87 & 745(10) & 222 & 423.44 &                                                                  $0.48$ & $0.443^{+0.004}_{-0.082}$ & $42.56^{+0.26}_{-0.37}$ & $40.24^{+0.27}_{-0.35}$     & Parkes I & [2][3] \\
010724 & $>30$ & 5.00 & >150 & 375 & 330.42 & 280.97 &                                                                  $0.33$ & $0.281^{+0.003}_{-0.072}$ & $>43.94$ & $>41.52$     & Parkes I & [4] \\
090625 & $1.14^{+0.42}_{-0.21}$ & $1.92^{+0.83}_{-0.77}$ & $2.19^{+2.10}_{-1.12}$ & 899.55(1) & 867.86 & 874.07 &       $0.98$ & $0.943^{+0.005}_{-0.094}$ & $43.87^{+0.28}_{-0.34}$ & $40.84^{+0.49}_{-2.84}$     & Parkes II & [5] \\
110220 & $1.30$ & $5.60^{0.10}_{-0.10}$ & $7.28^{+0.13}_{-0.13}$ & 944.38(5) & 909.61 & 920.26 &                        $1.03$ & $0.995^{+0.005}_{-0.094}$ & $43.94^{+0.26}_{-0.33}$ & $41.42^{+0.18}_{-0.19}$     & Parkes II & [6] \\
110523 & 0.60 & $1.73^{+0.17}_{-0.17}$ & 1.04 & 623.30(6) & 579.78 & 590.3 &                                            $0.67$ & $0.628^{+0.005}_{-0.089}$ & $43.12^{+0.25}_{-0.36}$ & $40.11^{+0.26}_{-0.32}$     & GBT & [7] \\
110627 & 0.40 & 1.40 & 0.56 & 723.0(3) & 675.54 & 689.43 &                                                              $0.77$ & $0.738^{+0.005}_{-0.091}$ & $43.08^{+0.27}_{-0.32}$ & $39.99^{+0.27}_{-0.31}$     & Parkes II & [6] \\
110703 & 0.50 & 4.30 & 2.15 & 1103.6(7) & 1061.27 & 1080.52 &                                                           $1.21$ & $1.176^{+0.006}_{-0.096}$ & $43.71^{+0.27}_{-0.33}$ & $40.97^{+0.32}_{-0.29}$     & Parkes II & [6] \\
120127 & 0.50 & 1.10 & 0.55 & 553.3(3) & 521.48 & 532.67 &                                                              $0.60$ & $0.564^{+0.005}_{-0.087}$ & $42.90^{+0.28}_{-0.36}$ & $39.77^{+0.25}_{-0.36}$     & Parkes II & [6] \\
121002 & $0.43^{+0.33}_{-0.06}$ & $5.44^{+3.50}_{-1.20}$ & $2.34^{+4.46}_{-0.77}$ & 1629.18(2) & 1554.91 & 1568.68 &    $1.78$ & $1.749^{+0.006}_{-0.098}$ & $44.19^{+0.36}_{-0.33}$ & $41.48^{+0.62}_{-0.60}$     & Parkes II & [5][8] \\
121102 & $0.40^{+0.40}_{-0.10}$ & $3.00^{+0.50}_{-0.50}$ & $1.20^{+1.60}_{-0.45}$ & 557(2) & 369 & 269.88 &             $0.32$ & $0.268^{+0.003}_{-0.070}$ & $42.07^{+0.49}_{-0.51}$ & $39.48^{+0.56}_{-0.82}$     & Arecibo & [9] \\
130626 & $0.74^{+0.49}_{-0.11}$ & $1.98^{+1.20}_{-0.44}$ & $1.47^{+2.45}_{-0.50}$ & 952.4(1) & 885.53 & 887.31 &        $0.99$ & $0.958^{+0.005}_{-0.094}$ & $43.77^{+0.47}_{-0.36}$ & $40.77^{+0.59}_{-0.64}$     & Parkes II & [5] \\
130628 & $1.91^{+0.29}_{-0.23}$ & $0.64^{+0.13}_{-0.13}$ & $1.22^{+0.47}_{-0.37}$ & 469.88(1) & 417.3 & 422.89 &        $0.48$ & $0.442^{+0.004}_{-0.082}$ & $43.28^{+0.20}_{-0.34}$ & $39.88^{+0.34}_{-0.54}$     & Parkes II & [5] \\
130729 & $0.22^{+0.17}_{-0.05}$ & $15.61^{+9.98}_{-6.27}$ & $3.43^{+6.55}_{-1.81}$ & 861(2) & 830 & 835.58 &            $0.93$ & $0.900^{+0.005}_{-0.093}$ & $43.12^{+0.40}_{-0.41}$ & $41.05^{+0.65}_{-3.05}$     & Parkes II & [5] \\
131104 & 1.12 & 2.08 & 2.33 & 779(1) & 707.9 & 558.8 &                                                                  $0.63$ & $0.593^{+0.005}_{-0.088}$ & $43.31^{+0.26}_{-0.35}$ & $40.46^{+0.24}_{-0.36}$     & Parkes II & [10] \\
140514 & $0.47^{+0.11}_{-0.08}$ & $2.80^{+3.50}_{-0.70}$ & $1.32^{+2.34}_{-0.50}$ & 562.7(6) & 527.8 & 538.53 &         $0.61$ & $0.571^{+0.005}_{-0.087}$ & $42.95^{+0.23}_{-0.38}$ & $40.23^{+0.62}_{-0.77}$     & Parkes II & [11] \\
150215 & $0.70^{+0.28}_{-0.01}$ & $2.80^{+1.20}_{-0.50}$ & $1.96^{+1.96}_{-0.37}$ & 1105.6(8) & 678.4 & 812.77 &        $0.91$ & $0.875^{+0.005}_{-0.093}$ & $43.66^{+0.23}_{-0.29}$ & $40.81^{+0.43}_{-0.37}$     & Parkes II & [14] \\
150418 & $2.19^{+0.60}_{-0.30}$ & $0.83^{+0.25}_{-0.25}$ & $1.82^{+1.20}_{0.72}$ & 776.2(5) & 587.7 & 450.66 &          $0.51$ & $0.473^{+0.004}_{-0.084}$ & $43.41^{+0.26}_{-0.33}$ & $40.10^{+0.46}_{-1.34}$     & Parkes II & [12] \\
150610 & $0.7^{+0.2}_{-0.2}$ & $2.0^{+1.0}_{-1.0}$ & $>1.3$ & 1593.9(6) & 1486.6 
& 1470.9 &               $1.66$ & $1.631^{+0.006}_{-0.097}$ & 
$44.18^{+0.29}_{-0.41}$ & $>41.04$     & Parkes II & [18] \\
150807 & $128.0^{+5.00}_{-5.00}$ & $0.35^{+0.05}_{-0.05}$ & $44.80^{+8.40}_{-7.90}$ & 266.5(1) & 196.5 & 241.43 &       $0.28$ & $0.235^{+0.003}_{-0.067}$ & $44.55^{+0.04}_{-0.46}$ & $40.83^{+0.33}_{-0.38}$     & Parkes II & [13] \\
151206 & $0.30^{+0.04}_{-0.04}$ & $3.0^{+0.6}_{-0.6}$ & $>0.9$ & 1909.8(6) & 
1666.4 & 1748.8 &            $2.00$ & $1.971^{+0.006}_{-0.097}$ & 
$44.02^{+0.26}_{-0.20}$ & $>41.07$     & Parkes II & [18] \\
151230 & $0.42^{+0.03}_{-0.04}$ & $4.4^{+0.5}_{-0.5}$ & $>1.9$ & 960.4(5) & 
912.47 & 922.6 &              $1.03$ & $0.997^{+0.005}_{-0.094}$ & 
$43.38^{+0.28}_{-0.15}$ & $>40.81$     & Parkes II & [18] \\
160102 & $0.5^{+0.1}_{-0.5}$ & $3.4^{+0.8}_{-0.8}$ & $>1.8$ & 2596.1(3) & 
2561.56 & 2574.3 &              $3.10$ & $3.076^{+0.005}_{-0.081}$ & 
$44.74^{+0.27}_{-0.28}$ & $>41.69$     & Parkes II & [18] \\
160317 & >3.0 & $21.00^{+7.00}_{-7.00}$ & >63.0 & 1165(11) & 845.4 & 770.38 &                                           $0.86$ & $0.828^{+0.005}_{-0.092}$ & $>44.17$ & $>42.14$     & UTMOST & [15] \\
160410 & >7.0 & $4.00^{+1.00}_{-1.00}$ & >28.0 & 278(3) & 220.3 & 221.29 &                                              $0.26$ & $0.211^{+0.003}_{-0.064}$ & $>43.06$ & $>40.51$     & UTMOST & [15] \\
160608 & >4.3 & $9.00^{+6.00}_{-6.00}$ & >38.7 & 682(7) & 443.7 & 371.69 &                                              $0.43$ & $0.384^{+0.004}_{-0.079}$ & $>43.44$ & $>41.24$     & UTMOST & [15] \\
170107 & 22.30 & 2.60 & 57.98 & 609.5(5) & 574.5 & 582.5 &                                                              $0.66$ & $0.620^{+0.005}_{-0.088}$ & $44.64^{+0.29}_{-0.34}$ & $41.86^{+0.28}_{-0.33}$     & ASKAP & [16] \\
170827 & $50.30$ & 0.40 & $19.87$ & 176.4 & 139.4 &149.4 &                                                              $0.18$ & $0.127^{+0.000}_{-0.050}$ & $43.38^{+0.33}_{-0.53}$ & $39.89^{+0.30}_{-0.46}$     & UTMOST & [17] \\
170922 & $2.30^{+0.50}_{-0.50}$ & 26.00 & 59.80 & 1111 & 1066 & 1078.11 &                                               $1.20$ & $1.173^{+0.006}_{-0.096}$ & $44.35^{+0.28}_{-0.33}$ & $42.43^{+0.27}_{-0.29}$     & UTMOST & [19]\\
171209 & 0.92 & 2.5 & 2.3 & 1458 & 1115 & 1223 &                                                                        $1.37$ & $1.339^{+0.006}_{-0.097}$ & $44.10^{+0.27}_{-0.29}$ & $41.17^{+0.25}_{-0.31}$     & Parkes II & [20]\\ 
180301 & 0.5 & 3.0 & 1.5 & 520 & 365 & 287 &                                                                            $0.33$ & $0.288^{+0.003}_{-0.072}$ & $42.19^{+0.30}_{-0.41}$ & $39.55^{+0.25}_{-0.37}$     & Parkes II & [21]\\
180309 & 20.8 & 0.576 & 11.98 & 263.47 & 218.78 & 233.5 &                                                               $0.27$ & $0.226^{+0.003}_{-0.066}$ & $43.56^{+0.28}_{-0.42}$ & $40.24^{+0.28}_{-0.42}$     & Parkes II & [22]\\
180311 & 0.2 & 12 & 2.4 & 1575.6 & 1530.3 & 1543.5 &                                                                    $1.75$ & $1.719^{+0.006}_{-0.098}$ & $43.70^{+0.26}_{-0.28}$ & $41.38^{+0.25}_{-0.29}$     & Parkes II & [23]\\
\hline
\end{tabular}
\begin{tablenotes}
\footnotesize
\item (a) peak flux density, (b) burst duration, (c) fluence of burst profile, 
	(d) observed dispersion measure,  (e) extragalactic DM computed using the 
	NE2001 model, and (f) extragalactic DM computed using the the YMW16 model,
	(g) maximum redshift inferred by extragalactic DM using the YMW16 model when assumed $\DMh=0$ and $\DMs=0$,
	(h) most probable redshift, (i) most probable isotropic luminosity, (j) most probable isotropic energy.
	\item For calculation of luminosity and energy, we assumed the FRB radiation 
		is isotropic with flat spectrum, and use 1 GHz as the reference value of 
		spectral bandwidth at rest frame of FRBs. The error bar is for 95\% 
		confidence level.
\item The references are, [1] \cite{Burke-Spolaor14ApJ}, [2] \cite{Keane11MN}, [3] \cite{Keane12MN}, 
        [4] \cite{Lorimer07Sci}, [5] \cite{Champion16MN}, [6] \cite{Thornton13Sci}, [7] \cite{Masui15Nat}, 
        [8] \cite{Thornton13PhDT}, [9] \cite{Spitler14ApJ}, [10] \cite{Ravi15ApJ}, [11] \cite{Petroff15MN}, 
	[12] \cite{Keane16Nat}, [13] \cite{Ravi16Sci}, [14] \cite{Petroff17MN}, [15] \cite{Caleb17MN}, 
	[16] \cite{Bannister17ApJ}, [17] \cite{Farah18MN}, 
	[18] \cite{Bhandari18MN}, [19] \cite{Farah17ATel},
	[20] \cite{Shannon17ATel}, [21] \cite{Price18ATel},
	[22] \cite{Oslowski18ATel1} and [23] \cite{Oslowski18ATel2}.
\end{tablenotes}
\label{tab:frbs}
\end{threeparttable}
\end{table}
\end{landscape}

\section{Derivation for marginalised likelihood}
\label{app:bayes}
Using random variable transformation, we can convert the PDF $f(\log L, r, \DMh, 
\DMs, \log\epsilon)$ to $f(\log S,\DMe, z, \DMs, \log\epsilon)$, i.e.  
\begin{equation}
\begin{aligned}
	& f(\log S,\DMe, z, \DMs, \log\epsilon) \\
	={} & |\MX{J}|\, f(\log L, r, \DMh, \DMs, \log\epsilon)\,,
\end{aligned}
\end{equation}
with the Jacobian determinant
\begin{equation}
\begin{split}
	\left|\MX{J}\right| &= \left|\begin{pmatrix}
	\frac {\partial \log L}{\partial \log S} & \frac{\partial \log L}{\partial \DMe} & \frac{\partial \log L}{\partial z} & \frac{\partial \log L}{\partial \DMs} & \frac{\partial \log L}{\partial \log\epsilon} \\
	\frac{\partial r}{\partial {\log S}} & \frac{\partial r} {\partial \DMe} & \frac{\partial r}{\partial z} & \frac{\partial r}{\partial \DMs} & \frac{\partial r}{\partial \log\epsilon} \\
	\frac{\partial \DMh}{\partial \log S} & \frac{\partial \DMh}{\partial \DMe} & \frac{\partial \DMh}{\partial z} & \frac{\partial \DMh}{\partial \DMs} & \frac{\partial \DMh}{\partial \log\epsilon} \\
	\frac{\partial \DMs}{\partial \log S} & \frac{\partial \DMs}{\partial \DMe} & \frac{\partial \DMs}{\partial z} & \frac{\partial \DMs}{\partial \DMs} & \frac{\partial \DMs}{\partial \log\epsilon} \\
	\frac{\partial \log\epsilon}{\partial \log S} & \frac{\partial \log\epsilon}{\partial \DMe} & \frac{\partial \log\epsilon}{\partial z} & \frac{\partial \log\epsilon}{\partial \DMs} & \frac{\partial \log\epsilon}{\partial \log\epsilon}
\end{pmatrix} \right|
	\label{eq:jocdet} \\
	&=\left|\begin{pmatrix}
		1 & 0 & \partial \log L/\partial z & 0 & 1 \\
		0 & 0 & c/H_0E(z) & 0 & 0\\
		0 & 1+z & \partial \DMh/\partial z & -1 & 0 \\
		0 & 0 & 0 & 1 & 0 \\
		0 & 0 & 0 & 0 & 1 \\
\end{pmatrix}\right| \\
&= \frac{c(1+z)}{H_0 E(z)}.
\end{split}
\end{equation}
Based on the modelling in \SEC{sec:meth}, we have
\begin{equation}
\begin{aligned}
	& f(\log S, \DMe, z, \DMs, \log\epsilon) \\
	={} &\phi(\log L) \fr(r) \fd(\DMh|z) \fs(\DMs) \fe(\log\epsilon) \frac{c(1+z)}{H_0 E(z)} \,,
\end{aligned}
\end{equation}
where $\DMh=(\DMe-\DMi)(1+z)-\DMs$.
The PDF of comoving distance $\fr(r) \propto r^2$ can be re-written as the 
PDF of redshift, i.e.
\begin{equation}
	\fz(z)=\fr(r) \frac{\D r}{\D z}\propto\frac{c\, r(z)^2}{H_0 E(z)}\,.
\end{equation}

To get the final likelihood, we need to marginalise the unknown information, 
i.e. $\DMs$, $\epsilon$, and $z$. The marginalisation of $\DMs$ leads to
\begin{equation}
\begin{aligned}
	f(\log S, \DMe, z, \log\epsilon) &= \phi(\log L)\, \fz(z)\, I(\DMe, z)\, \\
	& \quad \fe(\log\epsilon)\,(1+z)
\end{aligned}
\end{equation}
with
\begin{equation}
\begin{aligned}
	I(\DMe, z) &= \int_0^{\max(\DMs)}\fd(\DMh|z)\, \fs(\DMs)\, \D\DMs \, .
\end{aligned}
\label{eq:iz}
\end{equation}
The marginalisation for the beam response ($\epsilon$) gives
\begin{equation}
\begin{aligned}
	f(\log S, \DMe, z) &= \int_{-\log 2}^{0} f(\log S, \DMe, z, \log\epsilon)\, 
	\D\log\epsilon \\
	&= \fz(z)\, f(\DMe, z)\, I(\log L)\, ,
\end{aligned}
\end{equation}
with 
\begin{equation}
\begin{aligned}
	I(\log L) &= \int_{-\log2}^{0}\phi(\log L)\fe(\log\epsilon)\D\log\epsilon \\
	&=\frac{1}{\log 2}\left\{\Gamma\left[\alpha+1, \frac{L}{L^*}\right] - \Gamma\left[\alpha+1, \frac{2L}{L^*}\right]\right\}\, ,
\end{aligned}
\end{equation}
where $\Gamma$ is the incomplete \textsc{gamma} function.

Marginalisation of redshift ($z$) helps to get the final likelihood
\begin{equation}
	f(\log S, \DMe)=\frac{1}{N_{\rm f}}\int_0^{z_{\rm max}} f(\log S, \DMe,z) \, 
	\D z\,,
\end{equation}
where the maximal redshift ($z_{\rm max}$) in the upper limit of integration is 
computed by solving $\DMe-\DMi(z)=0$.
The normalisation factor $N_{\rm f}$ for the PDF is
\begin{equation}
	N_{\rm f}= \int_{\log S_{\rm min}}^{\infty} \D \log S\, \int\int\int f(\log S, \DMe, z, \log\epsilon)\, \D \DMe\, \D z\, \D\log\epsilon.
\end{equation}
After integrating $\DMe$ and $\log S$, one gets
\begin{equation}
	N_{\rm f}=\int_0^{z_{\rm max}} \fz(z)\,\D z \int \Gamma\left[\alpha+1, \frac{\max(L_{0}, 
	L_{\rm thre})}{\epsilon L^*}\right]\, \fe(\log\epsilon)\, \D\log\epsilon.
	\label{eq:normf}
\end{equation}
where $L_0$ is the lower cut-off of the luminosity function, $L_{\rm thre}\equiv4\pi 
r_{\rm L}^2 \Delta\nu_0 S_{\rm min}$ is the corresponding threshold luminosity 
for the survey sensitivity at the luminosity distance $r_{\rm L}$ with a perfect 
beam response $\epsilon=1$.

\section{Average electron density of galaxies}
\label{app:scaling}
We estimate the average electron density from the emission measure (EM), i.e.  
the integration of electron density variance along the line of sight  
$\mathrm{EM}\equiv\int n_{\rm e}^2\,\D r$.  EM can be derived from the $\Ha$ 
intensity \citep{Reynolds77ApJ}, i.e.
\begin{equation}
\mathrm{EM} = 2.75\left(\frac{T}{10^4\ \rm K}\right)^{0.9}\frac{I_{\rm 
\Ha}}{2.42\times10^{-7}}\, \mathrm{cm^{-6}\ pc}\,,
\label{eq:em}
\end{equation}
where $I_{\rm Ha}$ is the $\Ha$ intensity in units of $\mathrm{erg\, cm^{-2}\, 
s^{-1}\ sr^{-1}}$ and $T$ is the ionised gas temperature. $\Ha$ intensity 
($I_{\rm Ha}$) is calculated from the luminosity via
\begin{equation}
	I_{\rm \Ha}=\frac{L_{\Ha}}{4\pi r^2}=8\times10^{-5}\frac{L_{\Ha}}{10^{40}\, 
	\ergs} \left(\frac{r}{\rm kpc}\right)^{-2}\,,
	\label{eq:intapp}
\end{equation}
where $r=\eta R$ is the physical size of $\Ha$ emission region, in which
$R$ and $\eta$ are the radius of the galaxy and the filling factor respectively.
Combining \EQ{eq:intapp} and \EQ{eq:em}, we can derive the average electron 
density variance 
\begin{equation}
\left\langle n_e^2\right\rangle=1.0 \left(\frac{T}{10^{4}\, {\rm K}}\right)^{0.9}\left(\frac{L_{\Ha}}{10^{40}\, \ergs}\right)  
\left(\frac{R}{\rm kpc}\right)^{-3}\, \mathrm{cm}^{-6}.
\end{equation}
Then the average electron density of the whole galaxy is estimated using 
$\left\langle n_e\right\rangle\simeq\left\langle 
n_e^2\right\rangle^{1/2}\eta^3$, which leads to \EQ{eq:nea}.

\section{derivation of $\DMi$}
\label{sec:dmi}
Here, we calculate $\DMi$ in a rigorous fashion. To simplify the notations, 
we use natural units through out this section, where the speed of light $c=1$.

We assume a Robertson-Walker (RW) metric for the Universe that
$ds^2=-dt^2+a^2  \VEC{dx}^2$, where $a$ is the cosmic scale factor, and $\VEC{dx}$ is 
the spatial dual basis.
The local group velocity of radio wave propagating in the free electron gas is 
\citep{RL86}
\begin{equation}
	v_{\rm g}=\left(1+\frac{\beta n_{\rm e}}{\nu'^2}\right)^{-1}\,,
\end{equation}
where $\beta$ is the dispersion constant \citep{LK12HPA} and $\nu'$ is the radio 
wave frequency seen by local observer.
The corresponding propagation path associated with the flat-space RW metric is 
described by the differential equation
\begin{equation}
	\frac{\D r}{\D t}=\frac{1}{a} v_{\rm g}\,.
\end{equation}
As $\D z/\D t=(1+z)H_0E(z)$ and $1/a=1+z$, the solution to above differential 
equation gives
\begin{equation}
	r=\int_{z_1}^{z_2} \frac{1}{1+\beta n_{\rm e} \nu'^{-2}} \frac{1}{H_0
	E(z)} \, \D z,.
\end{equation}
The local electron density is $n_{\rm e}=\rho_{\rm c} \Omega_{\rm b} f_{\rm IGM} 
g(z) (1+z)^3 m_{\rm p}^{-1}$, where $f_{\mathrm{IGM}}$ is the cosmological 
baryon mass fraction in the IGM, the term $(1+z)^3$ comes from the Universe 
expansion, $\rho_{\rm c}=3 H_0^2/(8\pi G)$ is the Universe critical density, 
$m_{\rm p}$ is the proton mass. 

The frequency $\nu'$ from an emitter can be derived from the frequency ($\nu$) seen by 
the Earth observer, i.e.  $\nu'=\nu(1+z)$. Thus, \begin{equation}
	r(\nu)=\int_{z_1}^{z_2}\frac{1}{1+\beta \rho_{\rm c}\Omega_{\rm b} m_{\rm 
	p}^{-1} f_{\rm IGM} g(z) (1+z) \nu^{-2}} \frac{1}{H_0 E(z)} \, \D z\,.
	\label{eq:rtrw}
\end{equation}
At infinite frequency,
\begin{equation}
	r(\infty)=\int_{z_1}^{z_2}
	\frac{1}{H_0 E(z)} \, \D z\,.
	\label{eq:rtrwlight}
\end{equation}
By comparing \EQ{eq:rtrw} with (\ref{eq:rtrwlight}), the time delay is
\begin{equation}
	\Delta t=\frac{\beta}{\nu^2} \int \frac{\rho_{\rm c} \Omega_{\rm b} f_{\rm IGM}
	g(z)(1+z)}{m_{\rm p} H_0 E(z)}\, \D z\,,
	\label{eq:dmigmres}
\end{equation}
so that we have
 \begin{equation}
	 \DMi=\int \frac{\rho_{\rm c} \Omega_{\rm b} f_{\rm IGM} g(z)(1+z)}{m_{\rm p} H_0 E(z)}\, 
	 \D z\,,
 \end{equation}
which gives the same result as \citet{DZ14ApJ}.

\section{The most probable redshift, luminosity and energy}
\label{app:probl}

Using $\fd(\DMh)$, $\fs(\DMs)$ and $\fz(z)$, we can infer the most probable FRB redshift,
luminosity, and energy for each FRB individually. Similar methods have been applied to measure the pulsar 
distance \citep{VLM10MN, Verbiest12ApJ, IVC16}. We now treat the redshift PDF 
$\fz(z)$ as the prior. The posterior of redshift
given the extragalactic DM becomes
\begin{equation}
	f(z|\DMe)=\frac{1}{N_f}I(\DMe, z) \fz(z) (1+z) \,
	\label{eq:likz}
\end{equation}
where $I(\DMe, z)$ is the given in \EQ{eq:iz} and $N_f$ is the corresponding 
normalisation factor. The most probable redshift maximize the posterior, which 
leads to
\begin{equation}
	\hat{z}={\rm argmax}_{z} f(z|\DMe)\label{eq:zesti}\,.
\end{equation}
One can derive the most probable luminosity with the same method, of which
the posterior is \begin{equation}
\begin{aligned}
	f(\log L|\DMe, \log S)&=\frac{1}{N_f} \int \fz(z)\, (1+z) \, \D z\, \\
	& \quad \int f(\DMe, \log S|\log L,\log\epsilon, z) \\ & \quad \cdot 
	\fe(\log\epsilon)\, \D\log\epsilon\, .
	\label{eq:likL}
\end{aligned}
\end{equation}
Here we have assumed a uniform prior for $\log L$, and 
\begin{equation}
	f(\DMe, \log S|\log L,\log\epsilon,z)\propto \mathrm{exp}\left[-\frac{1}{2} \left( 
	\frac{\Delta S}{\sigma_{\rm S}}\right)^2\right] I(\DMe, z) \,,
	\label{eq:likl2}
\end{equation}
with \begin{equation}
		\Delta S= \frac{\epsilon 10^{\log L_\mathrm{iso}}}{4\pi r_{\rm L}^2 \Delta \nu_0}-10^{\log S}\,.
\end{equation}
The Gaussian likelihood is introduced to include the flux density measurement error
($\sigma_{\rm S}$). For those measurements without the corresponding errorbars, 
we take 30\% as the relative error and compute $\sigma_{\rm S}$. 

The intrinsic isotropic energy ($\mathcal{E}_\mathrm{iso}$) with a flat spectrum can be 
computed from the specific influence ($F$)
\begin{equation}
\mathcal{E}_\mathrm{iso}=\frac{F}{1+z}\, \Delta\nu_0\, 4\pi r_L^2\,.
\end{equation}
Thus, using a similar likelihood function compared to \EQ{eq:likL}, replacing $L$ with $\mathcal{E}$ and $S$ with $F$, the 
isotropic burst energy can be estimated by $F$. 

All the inferred parameter values are listed in \TAB{tab:frbs}.

\section{Posteriors of FRB luminosity functions}
\label{sec:postes}
The posterior distributions of Bayesian analysis are summarised here.

\begin{figure}
\centering
\subfloat[ETGs (NE2001)]
{\includegraphics[width=3.5in]{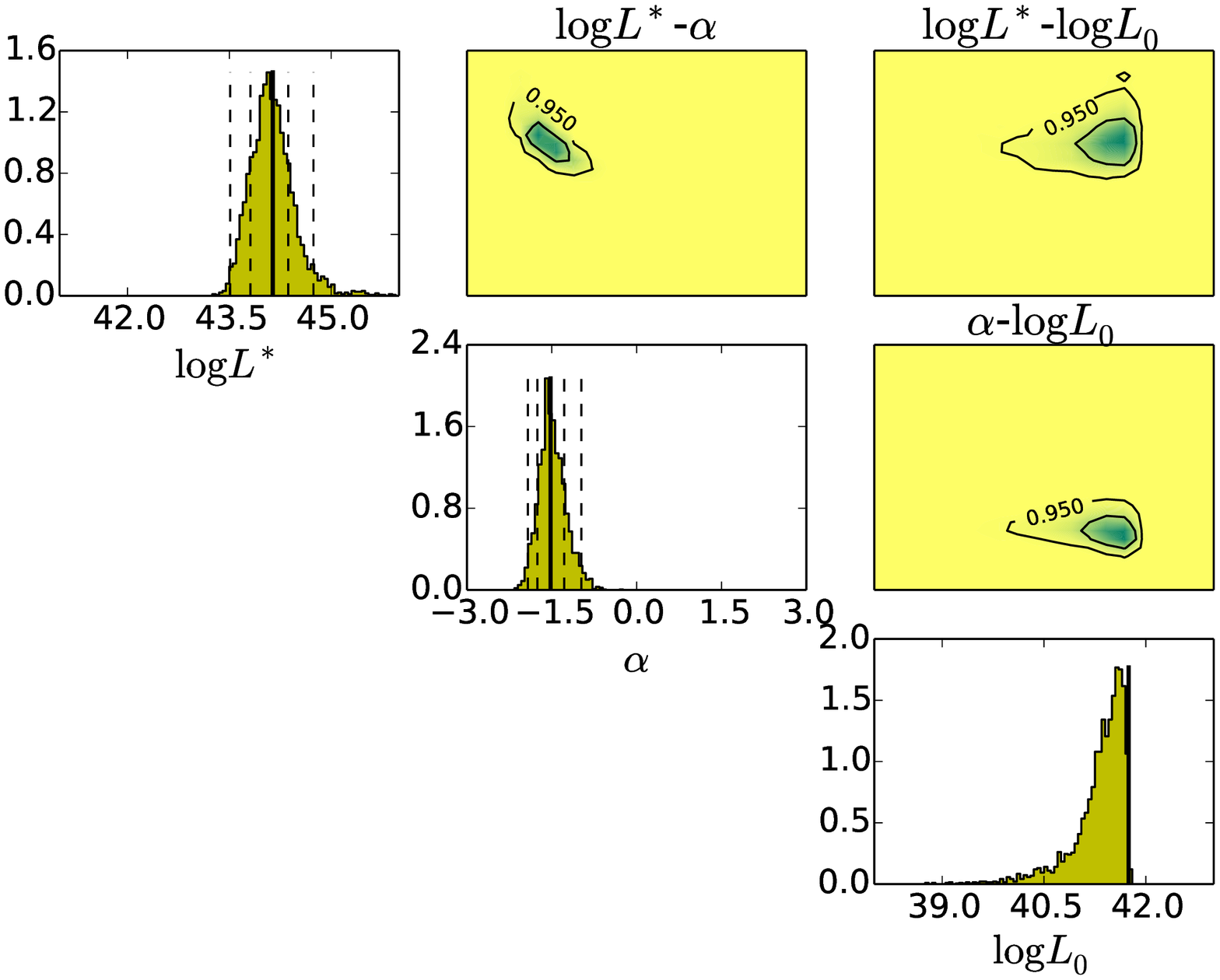}} \\
\subfloat[ETGs (YMW16)]
{\includegraphics[width=3.5in]{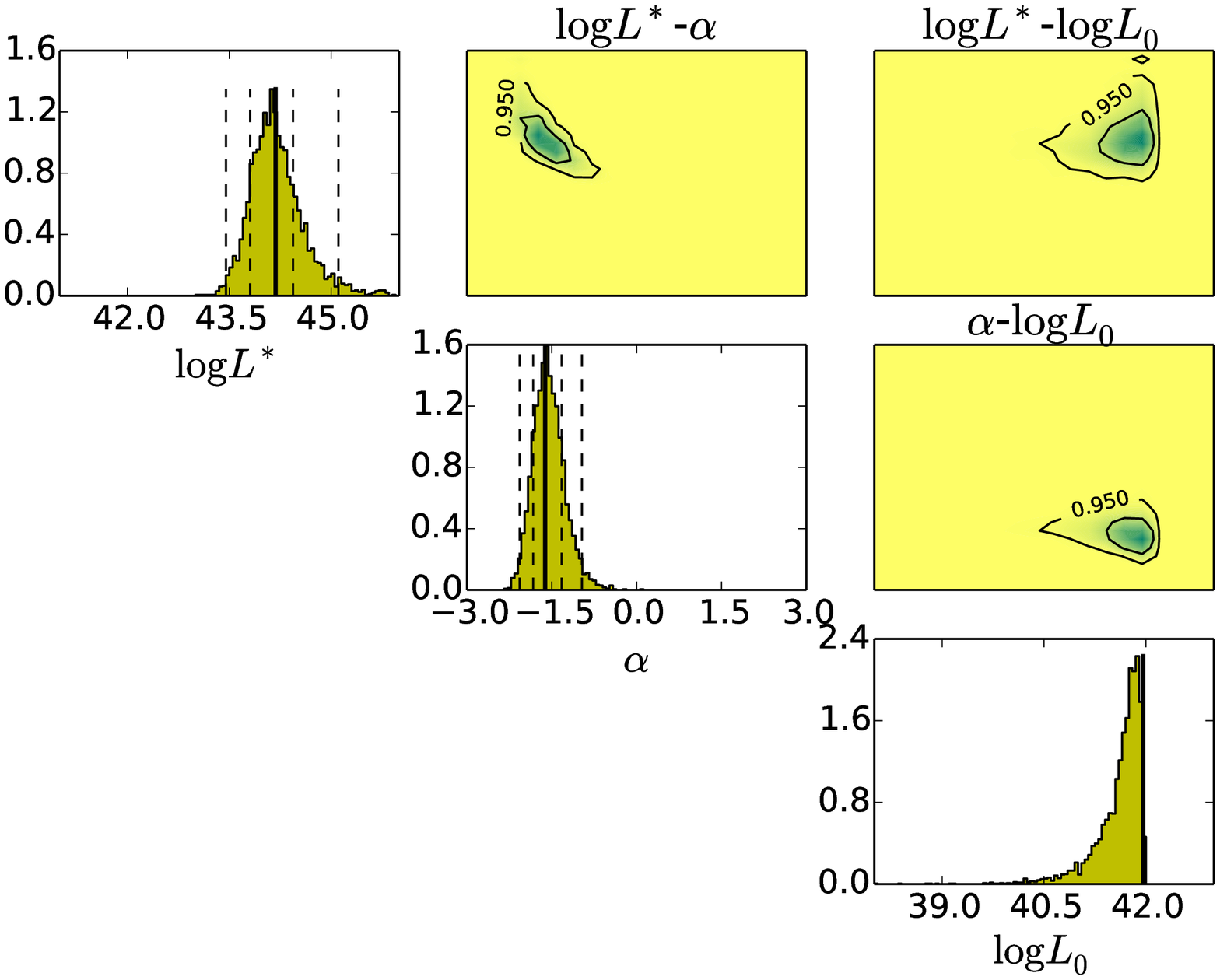}}
\caption{The marginalized posterior distribution of ETG luminosity function 
parameters. The diagonal histogram is the marginalised 
one-dimensional posterior distribution for each of the parameters. 
For $\log L^*$ and $\alpha$, the solid lines denote the most probable parameter value, while
the dashed lines indicate the 67\% and the 95\% confidence level. 
For $\log L_0$, the solid line denote the upper limit value with 95\% confidence level. 
The off-diagonal contour plots are for the marginalised two-dimensional posteriors, 
with parameters indicated in the title. The inner and outer black contours are for 67\% and 95\% 
confidence levels.}
\label{fig:post_etg}
\end{figure}

\begin{figure}
\centering
\subfloat[LTGs (NE2001)]
{\includegraphics[width=3.5in]{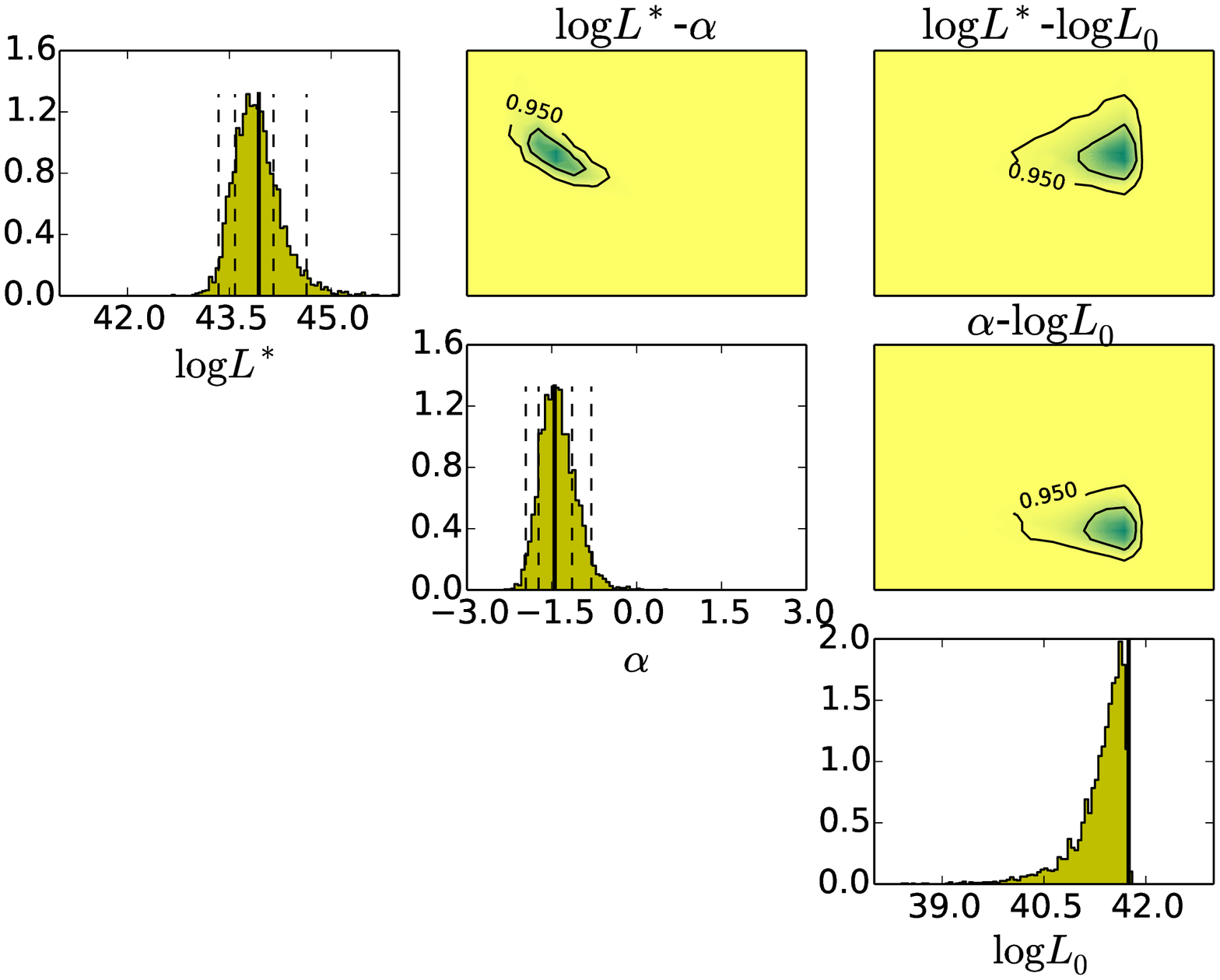}} \\
\subfloat[LTGs (YMW16)]
{\includegraphics[width=3.5in]{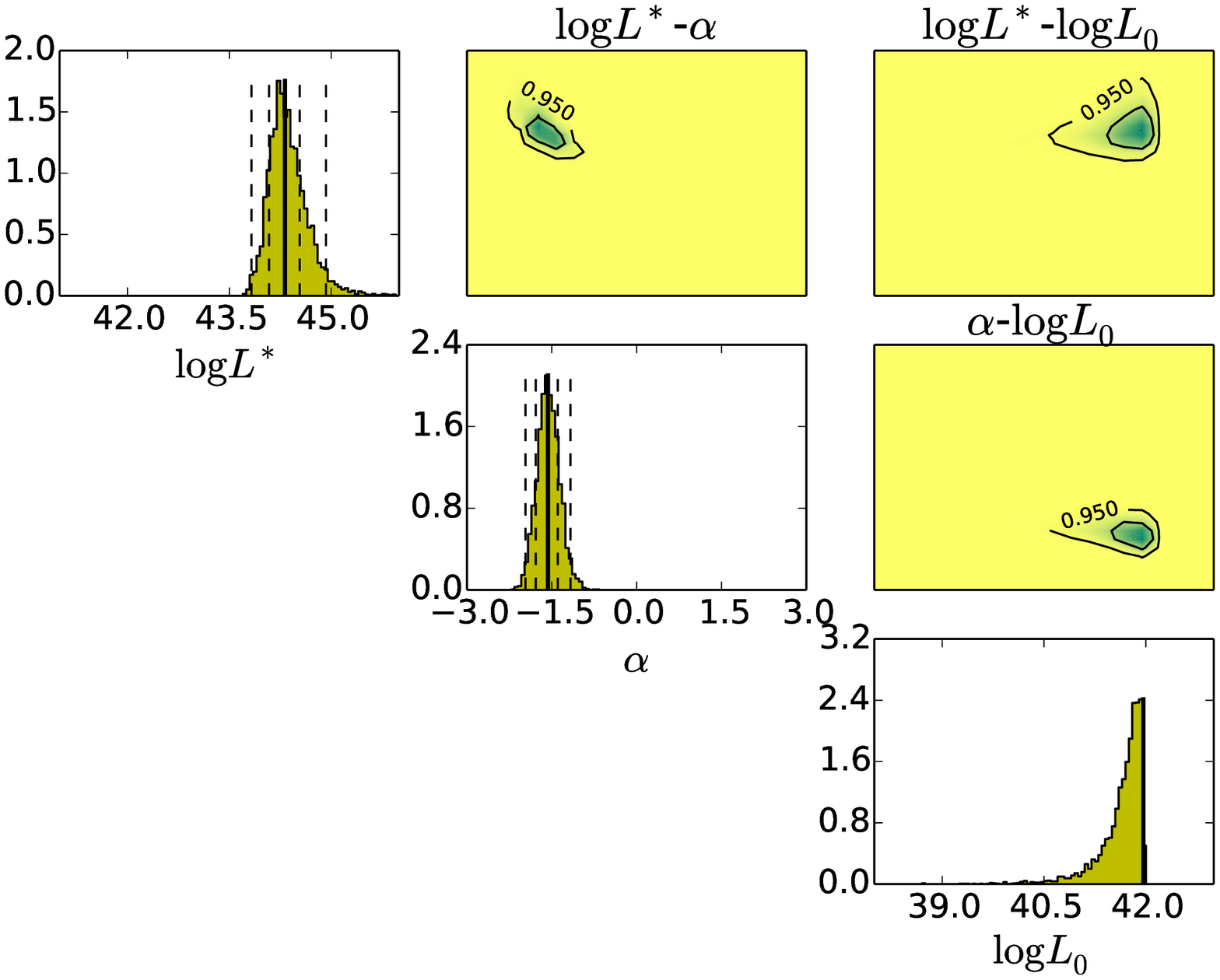}}
\caption{The marginalized posterior distribution of LTG luminosity function 
parameters. The plots details are the same as \FIG{fig:post_etg}.
}
\label{fig:post_ltg}
\end{figure}

\begin{figure}
\centering
\subfloat[ALGs (NE2001)]
{\includegraphics[width=3.5in]{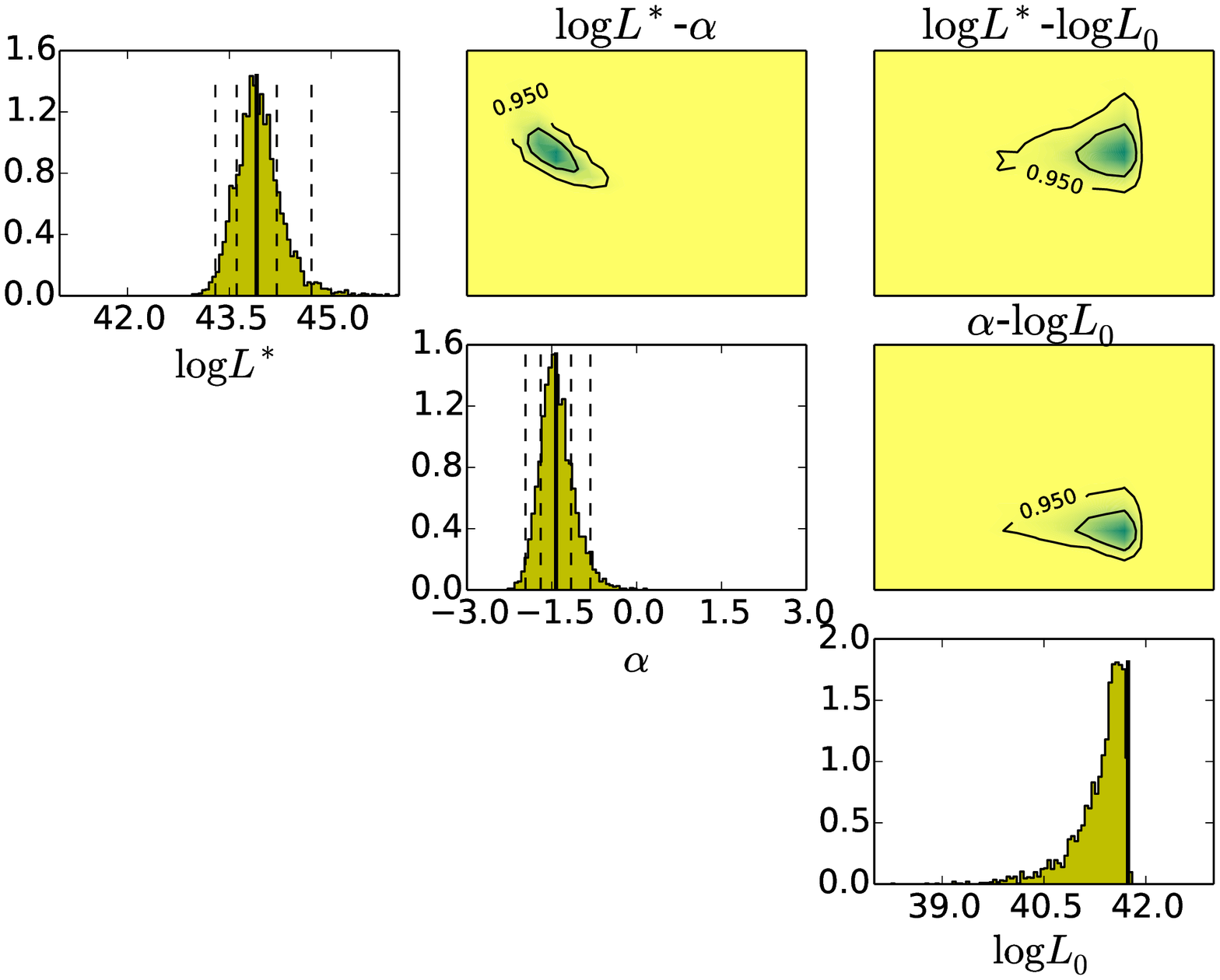}} \\
\subfloat[ALGs (YMW16)]
{\includegraphics[width=3.5in]{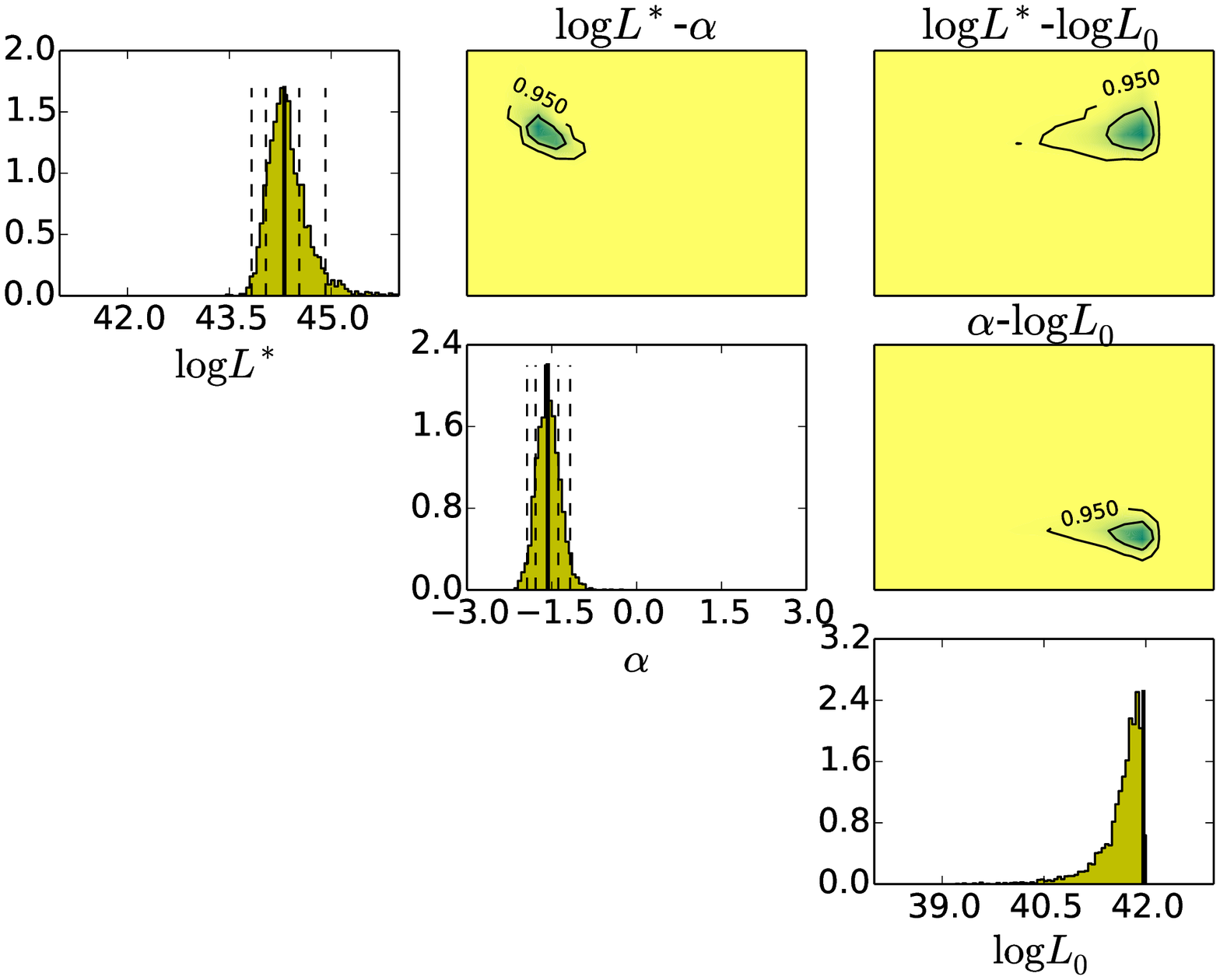}}
\caption{The marginalized posterior distribution of ALG luminosity function 
parameters. The plots details are the same as \FIG{fig:post_etg}.
}
\label{fig:post_alg}
\end{figure}

\bsp	
\label{lastpage}
\end{document}